\documentclass[fleqn,usenatbib,dvipdfmx]{mnras}
\DeclareRobustCommand{\VAN}[3]{#2}
\let\VANthebibliography\thebibliography
\def\thebibliography{\DeclareRobustCommand{\VAN}[3]{##3}\VANthebibliography}
\newcommand{\oneE}{1E~1547$-$5408}    
\newcommand{\NuS}{{\it NuSTAR}}
\newcommand{\Su}{{\it Suzaku}}
\newcommand{\etal}{{et al.}}
\usepackage{graphicx}	
\usepackage{amsmath}	
\usepackage{amssymb}	
\usepackage{booktabs}

\title{
A {\it NuSTAR} confirmation of the 36 ks hard X-ray pulse-phase modulation
in the magnetar 1E~1547.0$-$5408
}

\author[K. Makishima et al.]{
Kazuo Makishima,$^{1,2,3}$\thanks{E-mail: maxima@rikenjp (KM)}
Teruaki Enoto,$^{4}$
 Hiroki Yoneda,$^{3}$
and 
Hirokazu Odaka$^{2}$
\\
$^{1}$ Kavli Institute for the Physics and Mathematics of the Universe (WPI),
The University of Tokyo, 5-1-5 Kashiwa-no-ha,\\Kashiwa, Chiba, Japan 277-8683\\
$^{2}$Department of Physics, The University of Tokyo,
7-3-1 Hongo, Bunkyo-ku, Tokyo, Japan 113-0033\\
$^{3}$ High Energy Astrophysics Laboratory, and MAXI Team, RIKEN,
2-1 Hirosawa, Wako, Saitama, Japan 351-0198\\
$^{4}$ Extreme Natural Phenomena RIKEN Hakubi Research Team,
Cluster for Pioneering Research, RIKEN, 2-1 Hirosawa, \\
Wako, Saitama, Japan 351-0198
}
\date{Accepted 2021 January 15. Received 2021 January 15; in original form 2020 September 24}
\pubyear{2020}
\begin{document}
\label{firstpage}
\pagerange{\pageref{firstpage}--\pageref{lastpage}}
\maketitle

\begin{abstract}
This paper describes an analysis of the \NuS\ data of the fastest-rotating magnetar \oneE,
acquired in  2016 April for a time lapse of 151 ks.
The source was detected with a 1--60 keV flux of 
$1.7 \times 10^{-11}$ ergs s$^{-1}$ cm$^{-2}$,
and its pulsation at a period of $2.086710(5)$  sec.
In 8--25 keV,
the pulses were phase-modulated with a period of $T=36.0 \pm 2.3$ ks,
and an amplitude of $\sim 0.2$ sec.
This reconfirms  the \Su\ discovery of the same effect
at $T=36.0 ^{+4.5}_{-2.5} $ ks, 
made in the 2009 outburst.
These results strengthen the view 
derived from the \Su\ data, 
that this magnetar performs free precession
as a result of its axial deformation by $\sim 0.6 \times 10^{-4}$, 
possibly caused by internal toroidal magnetic fields reaching $\sim 10^{16}$ G.
Like in the \Su\ case, the modulation was not detected
in energies below $\sim 8$ keV.
Above 10 keV,  the pulse-phase behaviour, 
including the 36 ks modulation parameters,
exhibited complex energy dependences:
at $\sim 22$ keV, the modulation amplitude increased to $\sim 0.5$ sec,
and the modulation phase changed by $\sim 65^\circ$
over 10--27 keV,
followed by a phase reversal.
Although the pulse significance and pulsed fraction
were originally  very low in $>10$ keV, 
they both increased noticeably,
when  the arrival times of individual photons were corrected
for these systematic pulse-phase variations.
Possible origins of these complex phenomena are discussed,
in terms of  several physical processes 
that are specific to ultra-strong magnetic fields.

\end{abstract}


\section{Introduction}
\label{sec:intro}
A subclass of neutron stars (NSs) called magnetars 
 \citep[{\it e.g.},][]{Magnetar,Mereghetti08, Enoto17}
are thought to  have extremely strong magnetic fields (MFs)
with typical dipole field strengths of $B_{\rm d}=10^{14}-10^{15}$ G,
and emit X-rays by consuming their magnetic energies.
These NSs are considered to harbor, inside them, 
even higher toroidal MFs up to $B_{\rm t} \sim 10^{16}$ G.
Evidence for this inference includes strong differential rotation 
which is expected to take place in  the progenitor cores 
during their final collapse \citep[{\it e.g.},][]{toroidal_B},
and  the discovery of several objects 
which have rather low $B_{\rm d}$ 
and yet behave like magnetars \citep[{\it e.g.},][]{LowB_magnetar}.
However, the estimates of the toroidal MFs,
which are confined inside the stars, 
have obviously remained far more difficult than those of  $B_{\rm d}$,
which can be made  by measuring the pulse period  and its time derivative 
\citep{MaxReview}.

With  the Hard X-ray Detector \citep[HXD;][]{HXD1, HXD2} onboard \Su,
\citet{Makishima14} discovered a novel phenomenon
that  may  provide direct information on $B_{\rm t}$.
Specifically, through an observation of the magnetar 4U~0142+61 in 2009, 
they found that its 8.69 sec pulsation in the 15--40 keV energy range
repeats a slow {\it phase modulation},
by $\sim \pm 8$\% of a pulse cycle,
with a period of $T=55 \pm 4$ ks.
In a follow-up \Su\ observation of the same object in 2013 \citep{ Makishima19},
this phenomenon was reconfirmed at a consistent period of $T=54 \pm 3$ ks.
Furthermore, a \NuS\  data set of 4U~0142+61 
acquired in 2014 revealed the same modulation  at $T=54.8 \pm 5.3$ ks,
although the modulation amplitude was much smaller \citep{Makishima19}.

The $T \sim 55$ ks periodicity detected in the three observations 
of 4U~0142+61 has been interpreted in the following way \citep{Makishima14}.
Suppose that the star is axially deformed slightly by 
\begin{equation}
\epsilon \equiv (I_1 - I_3)/I_3~,
\label{eq:epsilon}
\end{equation}
where $\vec{I} = (I_1, I_2, I_3) $ is the moment of inertia
in the coordinates $(\hat{x}_1,\hat{x}_2,\hat{x}_3)$ fixed to the star,
with $\hat{x}_3$ the star's symmetry axis.
Then, the period $P_{\rm pr}$ of {\it free-precession} of the star
differs slightly  from its rotation period  $P_{\rm rot}$ around $\hat{x}_3$,
as $P_{\rm pr} = P_{\rm rot} (1+\epsilon)$.
If the star emits X-rays symmetrically around $\hat{x}_3$,
we would observe  regular X-ray pulses 
with a period of  $P_{\rm pr} $ (not $P_{\rm rot} $).
However, if the emission is asymmetric around $\hat{x}_3$,
the beat between $P_{\rm pr}$ and $P_{\rm rot}$ will 
modulate the observed pulse phase slowly with a  period of 
\begin{equation}
T=  P_{\rm pr}/ (\epsilon \cos \alpha) ~,
\label{eq:slip}
\end{equation}
which is called {\it slip period}.
Here, $\alpha$, or {\em wobbling angle},
is the angle  between $\hat{x}_3$ 
and the angular momentum $\vec{L}$ which is fixed to the inertial frame.
The modulation amplitude $A$, 
a fraction of $P$, depends positively on both  $\alpha$ 
and the emission asymmetry around $\hat{x}_3$.

By identifying  the observed 55 ks modulation period with this $T$, 
equation~(\ref{eq:slip}) indicates
\begin{equation}
\epsilon \cos \alpha =  P_{\rm pr}/ T = 1.6 \times 10^{-4}.
\label{eq:epsilon_0142}
\end{equation}
Assuming  $\cos \alpha \sim 1$,
we find $\epsilon \sim 10^{-4}$,
and postulating $\epsilon>0$  (a prolate shape), 
this deformation  may be explained as caused by  toroidal MFs hidden inside the NS.
In fact, from a theoretical calculation of magnetic deformation
of NSs as $\epsilon \sim 1 \times 10^{-4} (B_{\rm t}/10^{16}{\rm G})^2$ 
 \citep{Ioka+Sasaki04,Mastrano13},
the NS in 4U~0142+61 is inferred to  harbor
a toroidal MF of $B_{\rm t} \sim 10^{16}$ G \citep{Makishima14}.

A second example of the same phenomenon was discovered;
the fastest-rotating and highly variable magnetar \oneE,
with a pulse period of 2.07 sec.
As reported by \cite{Enoto10a},
this object was observed with  \Su\ on 2009 January 28 to 29,
just a week after the onset of its very bright outburst.
Analyzing the data obtained on this occasion,
\citet{Makishima16}, hereafter Paper I,
discovered that the  pulse phase in 15--40 keV 
is  periodically modulated with a period of 
\begin{equation}
T=36.0 ^{+ 4.5}_{-2.5}~{\rm ks}~,
\label{eq:36ks_Suzaku}
\end{equation}
and $A = 0.52 \pm 0.14$ sec which amounts to a quarter of $P_{\rm pr}$.
Then, equation (\ref{eq:slip}) yields $\epsilon \cos \alpha= 0.6 \times 10^{-4}$,
which is of the same order as equation (\ref{eq:epsilon_0142})  for 4U~0142+61.
Therefore, the two magnetars are suggested to have 
similar values of $B_{\rm t} \sim 10^{16}$ G.
Furthermore, of the two characteristic spectral components of magnetars
\citep{Kuiper06,Enoto10b},
{\it i.e.},  the Hard X-ray Component (HXC)
and the Soft X-ray Component (SXC),
only the HXC exhibited the phase modulation
in either object \citep{Makishima14,Makishima19}.

The results from \oneE\ are thus
generally consistent with those from 4U~0142+61,
and reinforce the scenario as explained above \citep{MaxReview}.
However,  the \Su\ pointing was a Target-of-Opportunity observation,
and covered only 2.4 cycles of $T=36$ ks.
Therefore, the phase modulation in the HXC pulses
could be some transient phenomena,
particularly those associated with the enhanced activity,
rather than a manifestation of persistent  free precession.
To reconfirm equation (\ref{eq:36ks_Suzaku}),
it is hence needed to observe \oneE\ in quiescence, 
and for a considerably longer period.
In fact, the object was observed again with \Su\ in 2010 August,
1.5 years after the first observation,
but the source was already rather faint,
 so its HXC was detected with the HXD only marginally \citep{Iwahashi13}.
This has motivated us to use \NuS\ instead.

\section{Observation }
\label{sec:observation}

Based on our proposal for the \NuS\
cycle-1 Guest Observation Program,
the present observation of \oneE\ (ObsID 30101035002) was 
conducted from 2016 April 23 UT 00:11:08 through April 24 UT 18:16:08,
for a total time lapse of 151 ks.
Through the standard data screening,
the two focal-plane instruments, FPMA and FPMB,
both achieved a net exposure of 83 ks.
By the present observation,
\oneE\ had already attained a relatively steady intensity,
which is  an order of magnitude lower
than in the 2009 \Su\ observation (\S~\ref{subsec:spectra}),
but still considerably higher than those in 2006
\citep{Kuiper12, Iwahashi13, Enoto17, CotiZelati20}.
The acquired data were already utilised by  \citet{CotiZelati20}, 
hereafter CZ20,
mainly with respect to the long-term source behaviour
through the outburst decay.

The on-source events from FPMA and FPMB were accumulated
within a circle of radius $60''$ around respective image centroids,
and the background events were taken from 
a source-free region on the respective images.
We applied dead-time corrections to the on-source data,
subtracted the background,
and performed vignetting corrections.
The object was detected at a  background-removed 
3--80 keV count rate of  $0.11$ c s$^{-1}$ for FPMA+FPMB,
in agreement with CZ20.
The arrival times of individual events were then  converted 
to those to be measured at the Solar-system barycentre,
utilizing  the source position of $(\alpha^{2000}, \delta^{2000})
= (15^{\rm h}50^{\rm m}54^{\rm s}.12,  -54^\circ 18'24''.11)$,
and  the spacecraft orbital information.

\section{Basic Data Analysis}
\label{sec:basic}
  
\subsection{Spectra}
\label{subsec:spectra}
\begin{figure}
\centerline{
\includegraphics[width=68mm]{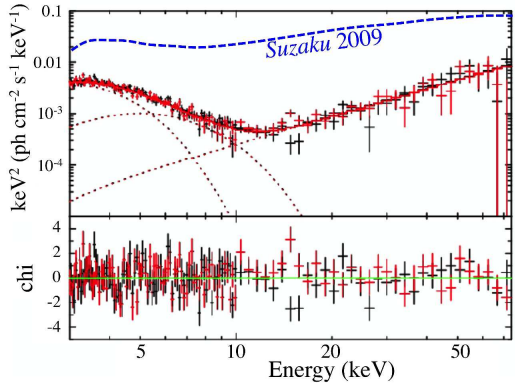}
}
\caption{Background-subtracted $\nu F \nu$ spectra of \oneE\
from FMPA (black) and FPMB (red).
They are jointly fitted by a model
consisting of two blackbodies for the SXC,
and a power-law for the HXC.
The \Su\ spectrum taken in the 2009 outburst
is superposed in a dahsed blue line,
also in the  $\nu F \nu$ form.}
\label{fig1:spec}
\end{figure}

Although the present paper aims at detailed timing studies of \oneE,
a brief spectral study would be necessary
in order to grasp approximate spectral characteristics, 
and to confirm the consistency with CZ20.
We hence created  background-subtracted FPMA and FPMB spectra separately,
and present them in Fig.~\ref{fig1:spec} in the $\nu F \nu$ form.
In agreement with Fig. 2 of CZ20, they exhibit a spectral shape 
characteristic of magnetars,
consisting of the SXC and the HXC
which are dominant in energies below and above $\sim 12$ keV, respectively.

We fitted the spectra jointly, 
using a model in which the HXC is expressed by a single power-law,
and the SXC with a sum of two blackbody components 
\citep[{\it e.g.},][]{Nakagawa09,Enoto11}.
This modeling is empirical rather than physical,
and may not be a unique description of  magnetar spectra in general,
or of the present spectra.
The fit was approximately successful 
with a reduced chi-square of $\chi^2/\nu = 253.1/211$, 
and yielded the HXC photon index as $\Gamma_{\rm h}  = 0.22  \pm 0.12$.
The two blackbody temperatures for the SXC
were obtained as $0.61 \pm 0.03$ keV and $1.2 \pm 0.1$ keV;
they are approximately in the 1:2 ratio
as  in many other magnetars \citep{Nakagawa09}.
These parameters are consistent with those of CZ20,
although they combined the 2016 data with another data set acquired in 2019.

Using the above model fitted to the \NuS\ spectra, 
we measured the  HXC flux as 
$F_{\rm h} = 5.6\pm 0.3$,
and the SXC flux after correcting for the absorption as
$F_{\rm s} = 11.1 \pm 0.1$,
both in 1--60 keV and in units of $10^{-12}$ erg s$^{-1}$ cm$^{-2}$.
The total 1--60 keV flux is hence $16.7 \times 10^{-12}$ ergs s$^{-1}$ cm$^{-2}$.
For comparison, the 2009 \Su\ data gave
$F_{\rm h} = 158.7$ and $F_{\rm s} =50.6$,
in the same units (Table 5 of \cite{Enoto17}).
Thus, from the 2009 observation,
the source became about an order of magnitude fainter,
with the HXC and SXC declining by a factor of 30 and 5, respectively.
The HXC/SXC flux ratio, 
previously $\xi \equiv F_{\rm h}/F_{\rm s} =3.1$ in 2009,
has decreased to $\xi =0.50 \pm 0.03$ accordingly.
This decrease in $\xi$ by a factor of $\sim 6$ is
somewhat larger  than the value of  2--3 reported by CZ20,
but this can be attributed to the different energy ranges 
utilised for the flux definition.
The two spectral measurements, in 2009 and 2019,
still satisfy the empirical scaling between $\xi$ 
and the characteristic age (common to the two data sets)
found by \citet{Enoto10b,Enoto17},
within the scatter by a factor of a few 
which is likely to be inherent to the scaling.

Compared to the \Su\ 2009 spectrum superposed in Fig.~\ref{fig1:spec},
the present spectra are characterised 
not only by the reduced fluxes,
but also by a clearer intensity minimum at $\sim 12$ keV
where the two spectral components cross over.
This can be attributed mainly to the harder HXC;
$\Gamma =1.3$ in 2009, $\Gamma \sim 1.1$ in 2010 \citep{Iwahashi13},
and $\Gamma =0.2$  in the present observation.
Therefore, another empirical scaling found by \cite{Enoto10b},
that older magnetars exhibit harder HXC (their Fig. 4a),
may be modified into a statement
that {\em less active magnetars show harder HXC}.

\begin{figure}
\centerline{
\includegraphics[width=75mm]{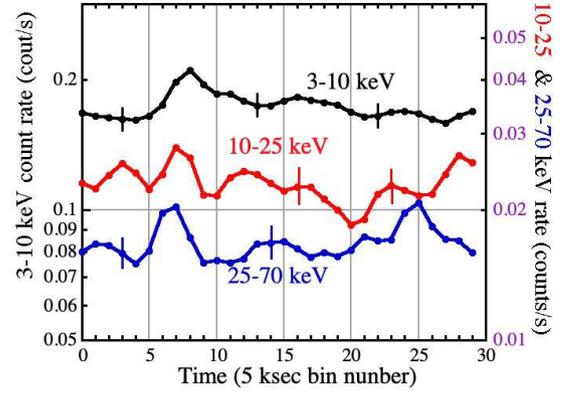}
}
\caption{Background-inclusive light curves of \oneE\ from FPMA+FPMB,
binned into 5 ks and corrected for dead time.
Black, red, and blue respectively represent 
3--10 keV (left ordinate), 10--25 keV (right ordinate),
and 25--70 keV (same) events.
The background rate is 
 $1.9 \times 10^{-3}$, $1.0 \times 10^{-3}$,
and  $1.1 \times 10^{-3}$ c~s$^{-1}$, 
in the 3--10, 10--25, and 25--70 keV bands, respectively.
}
\label{fig2:LTCV}
\end{figure}

\subsection{Light curves}
\label{subsec:light_curves}

Figure~\ref{fig2:LTCV} presents dead-time corrected 
but background-inclusive light curves of \oneE\  with  5 ks binning,
derived in 3--10 keV, 10--25 keV, and 25--70 keV.
Here and hereafter,
we use the  events from FPMA and FPMB co-added together,
because they provide a fully consistent pair of spectra.
As clear from Fig.~\ref{fig1:spec},
the first band represents the SXC,
whereas the latter two mainly the HXC.
Thus, the object was mildly variable 
with typical amplitudes of $\sim \pm 20 \%$,
on time-scales of several tens kiloseconds.
Over the bin number from 5 to 10,
the source brightened by  $\sim 30 \%$,
where the intensity maximum appears to propagate 
from the highest to the lowest bands in $\sim 10$ ks.
This suggests that the HXC variation causes the SXC to vary with some delay.
However, a standard cross-correlation analysis among the three light curves
did not yield meaningful results.

\subsection{The pulsation}
\label{sub:pulsations}

\subsubsection{Periodograms}
\label{subsub:PG}

To study the 2.07 sec source pulsation,
we utilise the  $Z^2$ technique.
Like in the conventional chi-square method,
we first fold the data, at a  trial period $P$,
into a folded  profile of $N_{\rm bin}$ bins,
$\{C_j(P);  j=0, 1, ., N_{\rm bin}-1\}$,
and express it as
\begin{equation}
C_j (P)= \sum_{k=0}^{N_{\rm bin}/2} \left\{
 a_k(P) \cos(\Omega k j) + b_k (P)  \sin (\Omega k j)
   \right\} ~.
   \label{eq:Fourier_tr}
\end{equation}
in a Fourier expansion.
Here,  $k=0, 1. .., N_{\rm bin}/2$ is the wave number, 
$\{a_k(P), b_k(P)\}$ are Fourier coefficients,
and  $\Omega\equiv 2 \pi/N_{\rm bin}$.
Then, the $k$-th Fourier power is obtained as
\[
Q_k(P)=  a_k(P)^2 + b_k(P)^2 ~.
\]
Summing up  $Q_k$
from the fundamental ($k=1$)  to a specified harmonic  $m$, 
and  normalising the result to the total photon counts $N_{\rm tot}$,
 the statistical quantity $Z_m^2$ is obtained as
\begin{equation}
Z_m^2(P) = \frac{1}{2N_{\rm tot} } \sum_{k=1}^{m} Q_k~,
\label{eq:Z2_definition}
\end{equation}
of which some properties  are given in Appendix A.
Together with $N_{\rm bin}=360$, we use  $m=4$,
 because the pulse profiles of magnetars
after the demodulation analysis
often exhibit three to four peaks 
per cycle \citep{Makishima19}.

\begin{figure}
\centerline{
\includegraphics[width=74mm]{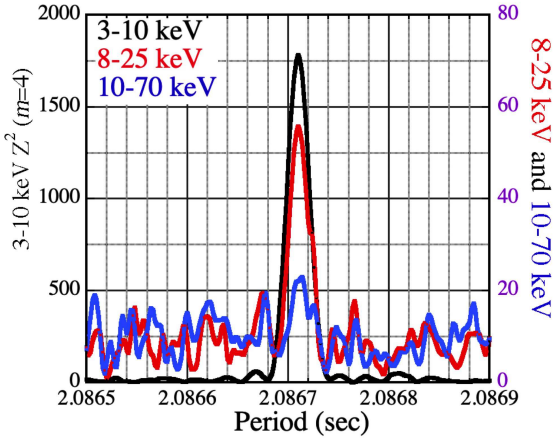}
}
\caption{Periodograms calculated from the 
background-inclusive FPMA+ FMPB data,
using the $Z^2$ method with $m=4$.
Black shows the result in the 3--10 keV band,
with the left ordinate.
Red and blue are those for the 8--25 keV and 10-70 keV bands,
respectively, with the right ordinate.}
\label{fig3:PG}
\end{figure}

By applying the $Z^2$ technique
to the background-inclusive data,
we calculated periodograms, {\it i.e.},
the values of $Z_4^2$ as a function of the trial period $P$.
The results are shown in Fig.~\ref{fig3:PG},
where the two harder bands used in Fig.~\ref{fig2:LTCV} 
were combined into one (blue; 10--70 keV)
to increase the statistics.
Instead, we have included a result from an intermediate band, 8--25 keV (red),
which plays an important role later on.
Thus, at least in the lower two energy bands,
the source pulsation is detected clearly, at a period of
\begin{equation}
P_0 = 2.086\; 710 \pm 0.000\; 005~{\rm sec}~,
\label{eq:P0}
\end{equation}
in full agreement with  CZ20.
The chance probability of this   peak is estimated as
$3.2 \times 10^{-9}$ in 8--25 keV, before counting trials.
Since the period range in Fig.~\ref{fig3:PG} is covered 
by $\sim 14$ independent Fourier wave numbers,
the probability  becomes $\sim 4.5 \times 10^{-8}$
when considering the frequency trial numbers.
This is still extremely low.

The  peak at $P_0$ is also visible in the 10--70 keV periodogram,
but its chance probability is $5 \times 10^{-3}$ and $\sim 0.07$
before/after considering the frequency trials, respectively,
and is comparable to several other peaks.
Therefore, we cannot claim the pulse detection in the 25--70 keV interval.

Compared with the period of  $2.072\; 135 \pm 0.000\; 005$ sec 
measured in 2009 January (\cite{Enoto10a};  Paper I),
equation~(\ref{eq:P0}) implies 
that the object has  been spinning down
with an average rate of $\dot{P}=6.4 \times 10^{-11}$ s s$^{-1}$,
which is somewhat larger than the long-term average of 
$3.4 \times 10^{-11}$ s s$^{-1}$ in 2 years from 2009 January 
as measured with  a dense sampling by \citet{Kuiper12}.
This is however not unusual,
because  $\dot{P}$ of this object and  of other magnetars
generally fluctuates considerably.

\subsubsection{Pulse profiles and the pulsed fraction}
\label{subsub:profiles}

\begin{figure}
\centerline{
\includegraphics[width=67mm]{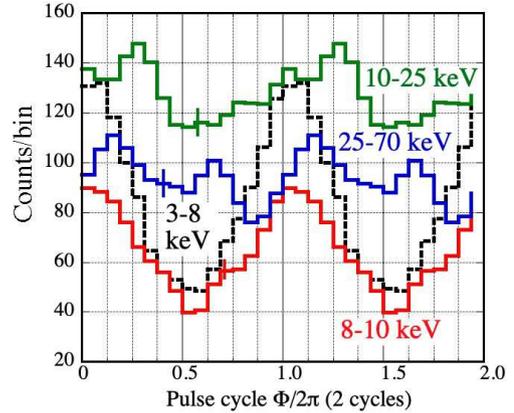}
}
\caption{Time-averaged pulse profiles 
in 3--8 keV (black dashed),
8--10 keV (red), 10--25 keV (green), and 25--70 keV (blue),
folded at $P_0$ and shown, with the running average,
as a function of the pulse cycle $\Phi/2\pi$.
The  ordinate is the background-inclusive  photon counts per bin,
where the 3--8 keV counts are scaled to 1/10.
}
\label{fig4:Pr}
\end{figure}

Figure~\ref{fig4:Pr} shows the folded pulse profiles
as a function of the pulse cycle, $\Phi/2\pi$.
Here, $\Phi$ is the pulse phase defined as
\begin{equation}
\Phi =2 \pi \; {\rm mod} (t/P_0 )~,
\label{eq:pulse_phase}
\end{equation}
where ``mod" means modulo,
and $t$ is the time as measured from  
a Mission Elapsed Time  of 199,067,862.632 sec.
The  three energy intervals utilised in Fig.~\ref{fig3:PG} 
have been re-arranged  into non-overlapping four energy bands.
Here and hereafter,  the pulse profiles are  presented 
after taking running averages, 
which combine three consecutive data bins
with weights of 0.25, 0.5, and 0.25.
As shown in Appendix B, 
this reduces the statistical errors,
associated with individual data bins, 
to 0.61 times the Poissonian values,
at the sacrifice of  independence
between adjacent data points.

In a similar way to equation (\ref{eq:Z2_definition}),
we define the pulsed fraction (PF) as
\begin{equation}
{\rm PF} = \frac{1}{N_{\rm tot}} 
\left[  \sum_{k=1}^{4} Q_k -\sum_{k=1}^{4} Q_k^{(n)} \right]^{1/2}~,
\label{eq:PF_definition}
\end{equation}
where $Q_{\rm k}$ again denotes the  Fourier power 
of the profile {\it before taking the running average},
and $Q_k^{(n)}$  is the expected  Poissionian-noise contribution.
This $Q_k^{(n)}$ was evaluated via a Monte-Carlo simulation,
wherein a constant pulse profile with the same average counts/bin
as the actual data was randomised, 
1000 times, to emulate the Poissonian noise.

For a purely sinusoidal waveform with $Q_k^{\rm (n)}\sim 0$,
the PF defined in this way coincides with a more simple-minded
definition of [pulse peak$-$pulse bottom]/average.
Like $Z_m^2$, this PF is independent of $N_{\rm bin}$ for  $N_{\rm bin} \gg m$,
but unlike $Z_m^2$, it is also independent of $N_{\rm tot}$
as long as  the pulse profile keeps a constant shape.
When $Q_k^{\rm (n)}$ can be ignored, we expect,
from equations~(\ref{eq:Z2_definition}) and (\ref{eq:PF_definition}),
\begin{equation}
Z_m^2 (P_0) \sim  \; \frac{N_{\rm tot}}{2} \times (\rm{PF})^2~.
\label{eq:PF_vs_Z2}
\end{equation}
Our PF definition is similar to that of CZ20,
except  some differences in the normalisation,
and in the harmonic number employed; while they used $m=2$,
we sum up to $m=4$ to be consistent with the $Z_m^2$ analysis.

Referring to Fig.~\ref{fig4:Pr},
the PF thus defined has been found to be 48.6\%, 35.4\%, 10.2\%, and 14.9\%,
 in 3--8, 8--10, 10--25, and  25--70 keV, respectively.
The results are consistent with Fig. 5 of CZ20,
considering the difference in the definition of PF.
These PF values in $<10$ keV from the present data,
and those measured by \cite{Kuiper12} through the outburst decline,
are higher than those derived during the 2009 outburst,
16--28\% in 1--50 keV  \citep{Enoto10a}, 
or before \citep{Halpern08}.
In addition, contrary to the observations in 2009
where the PF increased with energy \citep{Enoto10a},
the present PF drops markedly above $\sim 10$ keV,
which is also seen in Fig. 5 of CZ20.
Thus, the pulsed emission appears to have changed 
through the decay of the 2009 activity.

Figure~\ref{fig4:Pr}  also reveals some 
complex energy-dependent behaviour in the pulsation.
The  3--8 keV and 8--10 keV profiles
have a single peak at $\Phi/2\pi \sim 0$.
In 10--25 keV where the HCX becomes dominant, 
the pulse minimum  is still kept the same at $\Phi/2\pi \sim 0.5$,
but the peak at the pulse cycle $\sim 0$ is exceeded
by a stronger peak that appeared at $\Phi/2\pi \sim 0.3$.
In Fig.~5 of CZ20, this effect is observed as a 
phase difference, by $\sim 0.15$ pulse cycles,
between the 10--25 keV and $<10$ keV  profiles.
Our  25--70 keV profile shows two peaks,
and neither is in phase with those in the lower energies.

\begin{figure}
\centerline{
\includegraphics[width=80mm]{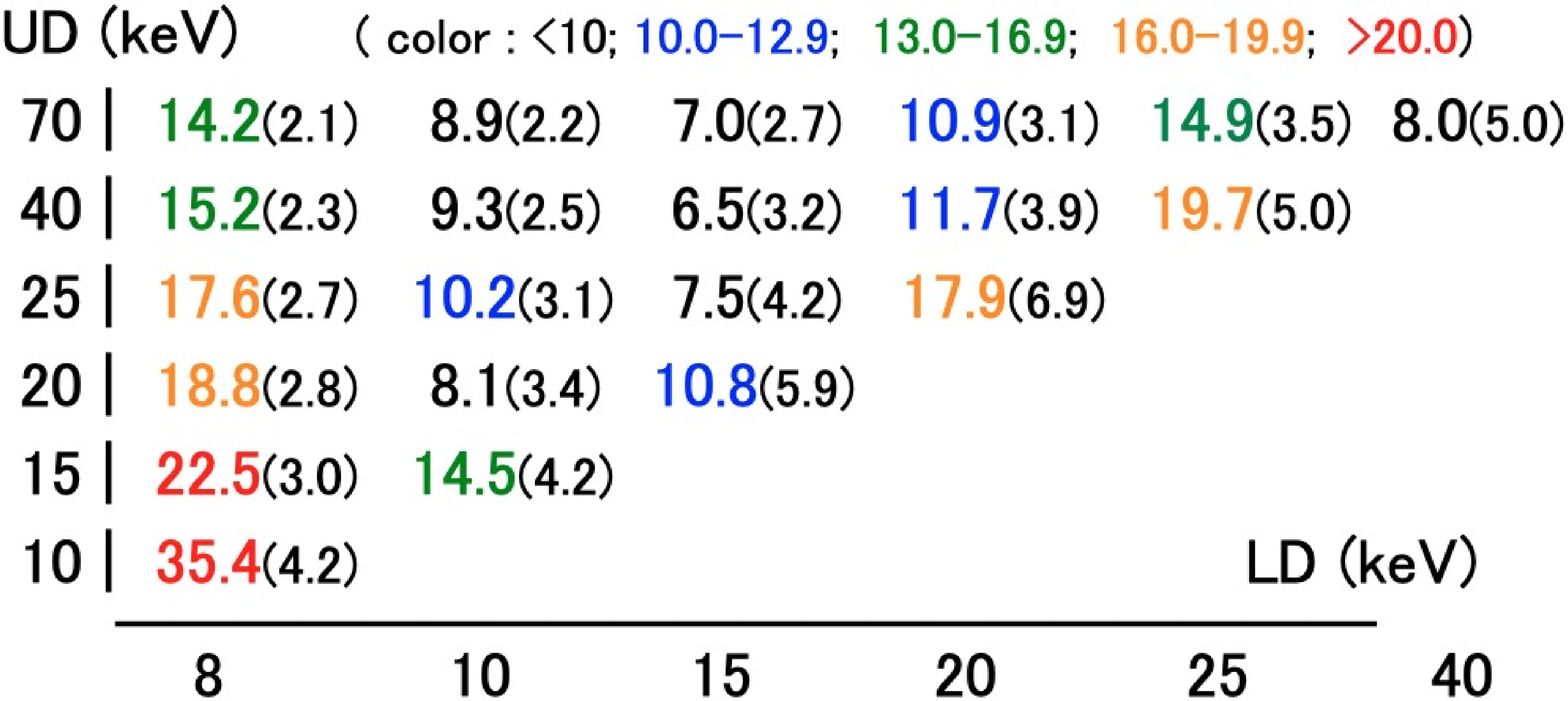}
}
\caption{The PF (see text) in percent, 
shown as a function of 
the LD (abscissa) and the UD (ordinate)
 in a matrix form.
The numbers in parenthesis give 1-sigma statistical errors,
 estimated by the Monte-Carlo simulation that was
 employed to calculate $\sum Q_k^{(n)}$  in equation~(\ref{eq:PF_definition}).
}
\label{fig5:LDUD}
\end{figure}

To more systematically study the PF behaviour,
we fixed $P=P_0$, and produced pulse profiles at  finer energy bands,
by changing  the lower- and upper-threshold energies,
abbreviated as  LD and UD, respectively.
The derived PF values 
are given in Fig.~\ref{fig5:LDUD},
on the LD vs. UD plane in a matrix form.
It reconfirms the implication of Fig.~\ref{fig4:Pr},
that the PF is rather low at intermediate energies.
The figure further indicates a hint of PF increase 
towards the diagonal matrix line.
This suggests that the  pulse coherence degrades
when the data are summed over wider energy ranges,
possibly due to some energy dependences 
in the phases and/or shapes of the HXC pulses
as pointed out by CZ20.

\begin{figure*}
\centerline{
\includegraphics[height=55mm]{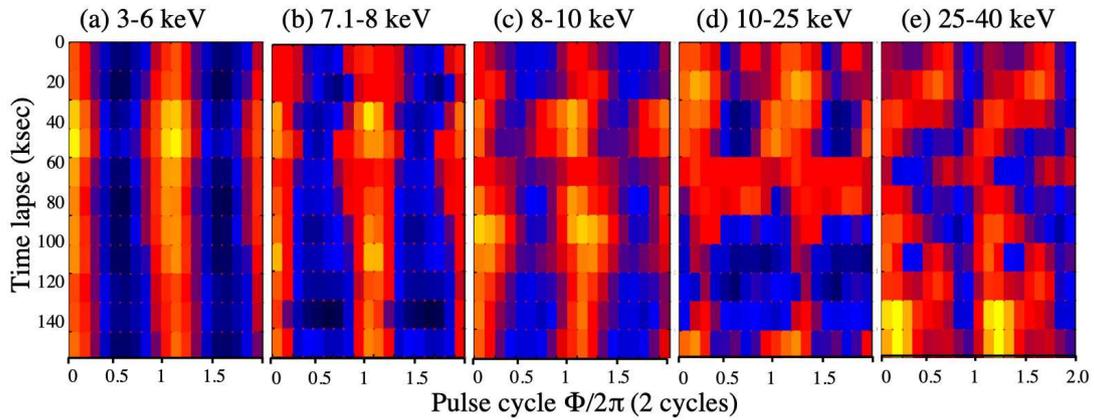}
}
\caption{Dynamic pulse profiles (see text) in 5 energy bands.
The color indicates the background-inclusive intensity,
increasing from black, blue, red, to yellow.
The abscissa is the pulse cycle $\Phi/2\pi$  (2 cycles),
whereas the ordinate is the time lapse in ks
running downwards from the start to the end of the observation.
The exposure correction and the running averages are
applied in the $t$ and $\Phi$ directions, respectively.}
\label{fig6:DPP}
\end{figure*}

\subsubsection{Dynamic pulse profiles}
\label{subsub:dynamic_profiles}

Although Fig.~\ref{fig4:Pr} averages over the total data length,
the HXC pulse profile of a magnetar is known to vary
on time-scales of hours \citep[{\it e.g.},][]{Enoto11},
partially coupled with free precession \citep{Makishima19}, 
as well as in months through an outburst decay \citep{Kuiper12}.
The PF would then degrade in some energy ranges,
if such profile/phase changes are enhanced therein.
To grasp this possibility,
we use a plot to be called a {\it dynamic pulse profile};
it is a two-dimensional color map,
where the abscissa  is the pulse cycle, 
the ordinate shows the time lapse  $t$,
and the color represents the X-ray intensity  at $(\Phi/2\pi,t)$.
Using the Good Time Interval information,
we correct the map  for exposure in the $t$ dimension,
but not in the $\Phi$ direction where the exposure is uniform to within $\sim 2\%$.
In the $\Phi$ direction,  we instead apply the running average,
to make the results consistent with time-integrated pulse profiles
such as in Fig.~\ref{fig4:Pr}.
The obtained results are shown in Fig.~\ref{fig6:DPP}.

The 3--6 keV dynamic pulse profile in panel (a),
representing the SXC pulse behaviour,
is dominated by a straight vertical ridge formed by the pulse peak,
at a pulse cycle 0--0.1 in agreement 
with the 3--8 keV profile in Fig.~\ref{fig4:Pr}.
Except some variations in the ridge height
due to the source variability (Fig.~\ref{fig2:LTCV}),
the pulse phase and shape are rather constant.
The pulse behaviour looks still similar in panel (b),
where we narrowed the energy interval to 7.1--8.0 keV
to purposely reduce the signal statistics.
In contrast, the 8--10 keV result in panel (c) 
reveals larger pulse-phase fluctuations with time.
This is likely to be intrinsic,
because the energy ranges of (b) and (c) were selected 
to have similar (within 10\%) pulse significances
in terms of the pulse amplitude relative to the Poisson noise.
In the two higher energy intervals,
these fluctuations further increase,
though due partially to increasing Poissonian errors.
In Fig.~\ref{fig6:DPP}e, an interpulse emerges 
at a pulse cycle $\sim 0.6$, 
in agreement with Fig.~\ref{fig4:Pr} (blue).

Figure~\ref{fig6:DPP} thus suggests
that the  pulse phase/shape in $\gtrsim 10$ keV fluctuates with time.
This could be responsible for the pulse-coherence degradation
 seen in these energies (Fig.~\ref{fig5:LDUD}).
Furthermore, if the fluctuation is energy dependent,
the degradation would be enhanced
when the data are summed over wider energy intervals.

\section{Detailed Pulsation Analysis}
\label{sec:detailed_pulsation}

If the time- and energy-dependent variations
in the HXC pulse properties (\S~\ref{sub:pulsations}) are of sporadic nature,
there would be little hope to recover the pulse coherence.
However, if they are of systematic nature,
and quantified as functions of time and energy,
we may  recover the pulse coherence 
through corrections for the systematic effects.
This is similar to the case of a pulsar in a binary,
where the pulsation generally becomes much more significant
when the arrival times of individual photons 
are corrected for the pulsar's binary motion.
This section is devoted to such attempts.

\subsection{Demodulation analysis of the 8--25 keV data}
\label{subsec:demod_8-25keV}

Putting off the energy dependent corrections to 
\S~\ref{subsec:demod_various_energies},
we  first consider the time dependence,
assuming that the HXC pulse-phase variation is periodic
as actually observed in 4U~0142+61 (\S~\ref{sec:intro}),
and  \oneE\ in 2009 (Paper I).
To search for any preferred periodicity (including 36 ks in particular)
in the pulse-phase fluctuation in the present data,
we perform a so-called demodulation analysis 
(\cite{Makishima14}; Paper I; \cite{Makishima19}).
We start this analysis with the 8--25 keV energy band,
where the pulse is suggested to fluctuate with time 
as in  panels (c) and (d) of Fig.~\ref{fig6:DPP},
but the PF is  still not too low,  17.6\% (Fig.~\ref{fig5:LDUD}).

\begin{figure}
\begin{center}
\includegraphics[width=77mm]{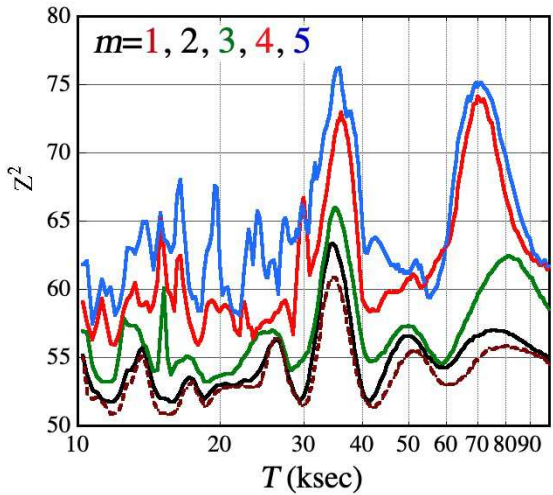}
\end{center}
\caption{The DeMDs for the 8--25 keV  data.
The abscissa shows the assumed modulation period $T$,
wheres the ordinate represents the maximum $Z_m^2$ 
when $A$, $\psi$, and $P$ are optimized at each $T$.
From bottom to top, the curves represent $m=1$ (dashed brown),
$m=2$ (black), $m=3$ (green), $m=4$ (red), and $m=5$ (blue).
}
\label{fig7:Tscan_8-25keV}
\end{figure}

The demodulation analysis postulates
that  the  arrival time  $t$ of each pulse from the pulsar 
deviates  from an exact periodicity by an amount 
\begin{equation}
\delta t = A \sin \left\{ \Psi (t) -\psi \right\}~,
\label{eq:modulation}
\end{equation}
where 
\begin{equation}
\Psi (t)= 2 \pi \; {\rm mod} (t/T)
\label{eq:modulation_phase}
\end{equation}
is the modulation phase defined in a similar way to 
$\Phi$ of equation~(\ref{eq:pulse_phase}),
$T \gg P$ and $A< P$  describe the period and amplitude of 
the assumed pulse-phase modulation, respectively,
and $\psi$ is  the initial modulation phase (with the sign reversed).
Then, by shifting the arrival times $t$ of individual photons
(instead of those of individual pulses) back by $-\delta t$,
we re-calculate the pulse periodogram, 
and search for a triplet $(T, A, \psi)$ 
that maximizes $Z_{\rm m}^2 (P)$ at $P \sim P_0$ of equation~(\ref{eq:P0}).
In practice, $T$ was varied from 10  ks to 100 ks with a step of 0.2--1.0 ks,
$A$  from 0 to 0.6 sec ($\gtrsim P/4$) as in Paper I, with a step of mainly 0.01 sec,
and $\psi$ from 0 to $360^\circ$ with a $10^\circ$ step.
For each triplet, the  pulse period $P$ was scanned 
over the error range of equation~(\ref{eq:P0})
with a 2 $\mu$ sec step.

Figure~\ref{fig7:Tscan_8-25keV}, 
which we call a {\it demodulation diagram} (hereafter DeMD),
summarises the results of this analysis
using the 8--25 keV data.
The abscissa represents $T$,
and the ordinate the maximum $Z_m^2$ obtained at each $T$,
when $A$, $\psi$, and $P$ are varied within the respective scan ranges.
The  results  with $m=1$ to 5
all reveal prominent peaks at  $T\sim 36$ ks.
Their widths of $\Delta T \sim 8$ ks, 
which satisfy $\Delta T/T \sim T/151$,
are consistent with being  determined by the  data length of 151 ks.
For $m=4$ and $m=5$,
an additional peak is seen at $T \sim 72$ ks,
which is just twice that of the main peak.

\begin{figure*}
\centerline{
\includegraphics[width=105mm]{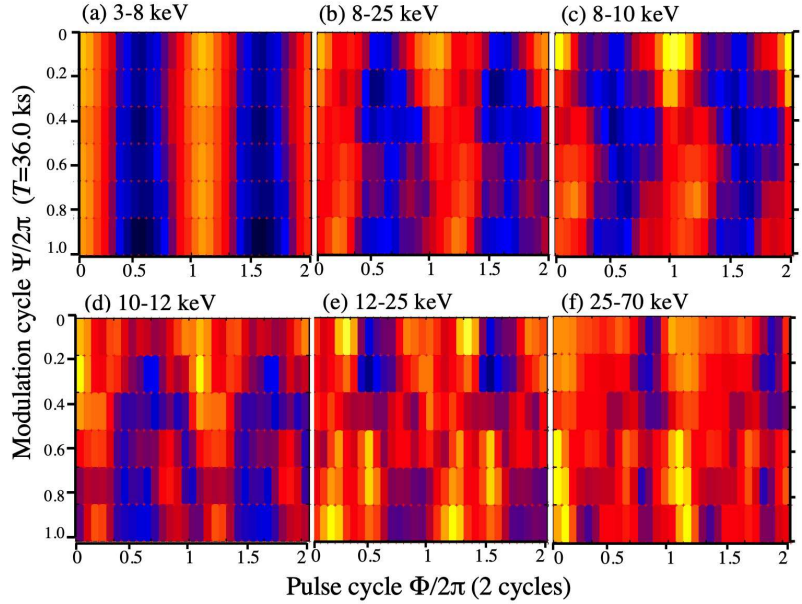}
}
\caption{Double-folded maps in 6 energy bands,
each shown on a plane formed by  the pulse cycle $\Psi/2\pi$
and the cycle $\Psi/2\pi$ of the 36 ks modulation periodicity.
The color coding, 
the exposure correction,
and the running average
are the same as  in Fig.~\ref{fig6:DPP}.
}
\label{fig8:dblfld_6bands}
\end{figure*}

When selecting $m=4$ as a representative,
the peak at 36 ks  is characterised by
\begin{equation}
T = 36.0 \pm 2.3 ~{\rm ks}~,~~
A = 0.19 \pm 0.06~{\rm s}~, ~~ 
\psi=150^\circ \pm 25^\circ~
 \label{eq:demodpara_8-25keV}
\end{equation}
and 
\begin{equation}
Z_4^2 =72. 95~~,~~~\delta Z_4^2 =17.23
 \label{eq:demodpara2_8-25keV}
 \end{equation}
together with $P$ of equation~(\ref{eq:P0}).
Here,  $\delta Z_m^2$ is the increment in $Z_m^2$ 
relative to the value before the demodulation,
and provides a measure of the pulse-significance increase (Appendix A).
These parameters obtained in the demodulation analysis
are summarised in Table~\ref{tbl:Z2_summary},
together with those obtained later in other energy intervals.
As given there, the demodulation has increased the 8--25 keV PF
from 17.6\% to 20.7\%, reflecting equation~(\ref{eq:PF_vs_Z2}).
Hereafter, we keep using  $m=4$,
because $\delta Z_4^2$ in  equation~(\ref{eq:demodpara2_8-25keV})
is considerably larger than 
$\delta Z_1^2 =10.22$, 
$\delta Z_2^2 =11.94$, and 
$\delta Z_3^2 =12.58$.
(Although $\delta Z_5^2=19.89$ is still larger,
the DeMD becomes noisier.)

As explained in Appendix C, 
the errors associated with the demodulation parameters
have been estimated as the range 
where the $Z_4^2$ values stay between the maximum
and the maximum minus $4.72$.
This updates the more conventional method of error estimation,
adopted so far by \citet{Makishima14}, Paper I, and \citet{Makishima19}.

The value of $T$ of the DeMD peak in Fig.~\ref{fig7:Tscan_8-25keV}
(equation~\ref{eq:demodpara_8-25keV}) is consistent  
with equation~(\ref{eq:36ks_Suzaku}) derived with \Su.
Moreover, as detailed  in Appendix D,  
we find a chance probability of $\sim  0.5\%$
for a $Z_4^2$ peak higher than  equation (\ref{eq:demodpara2_8-25keV})
to appear at $T$ that is consistent with equation~(\ref{eq:36ks_Suzaku}).
We hence conclude
that the 36 ks pulse-phase modulation detected with \Su\
has been reconfirmed with \NuS,
at least in the 8--25 keV energy band.
Although  $A$ is much  smaller 
than the \Su\ result of  $A=0.52 \pm 0.14$ sec (Paper I),
this would not be a problem,
because marked changes in $A$ have been observed
from 4U~0142+61 \citep{Makishima19}.

To visualise the 36 ks pulse-phase modulation,
Fig.~\ref{fig8:dblfld_6bands} presents
another form of two-dimensional X-ray count-rate map,
which we call a {\it double-folded map}.
It is obtained by sorting the photons into two dimensions,
where the abscissa $\Phi/2\pi$ again specifies the pulse cycle,
and the ordinate $\Psi/2\pi$ 
the  modulation cycle modulo $T$.
As before, the exposure correction is applied only 
in the $\Psi$-dimension,
and the running average only  in the $\Phi$-dimension.
It is thus similar to  Fig.~\ref{fig6:DPP}, but differs in the  ordinate.
The 3--8 keV map in panel (a) 
shows a straight pulse ridge running  at  $\Phi/2\pi \sim 0.05$,
as expected from panels (a) and (b) of  Fig.~\ref{fig6:DPP}.
In contrast, in panel (b) using the  8--25 keV photons,
the ridge is observed to wiggle laterally
by $\pm 0.1$ pulse cycles,
in agreement with equation~(\ref{eq:demodpara_8-25keV}).
This is the 36 ks pulse-phase modulation in 8--25 keV.
The other panels of Fig.~\ref{fig8:dblfld_6bands} are utilised later.

\begin{figure}
\begin{center}
  \includegraphics[width=60mm]{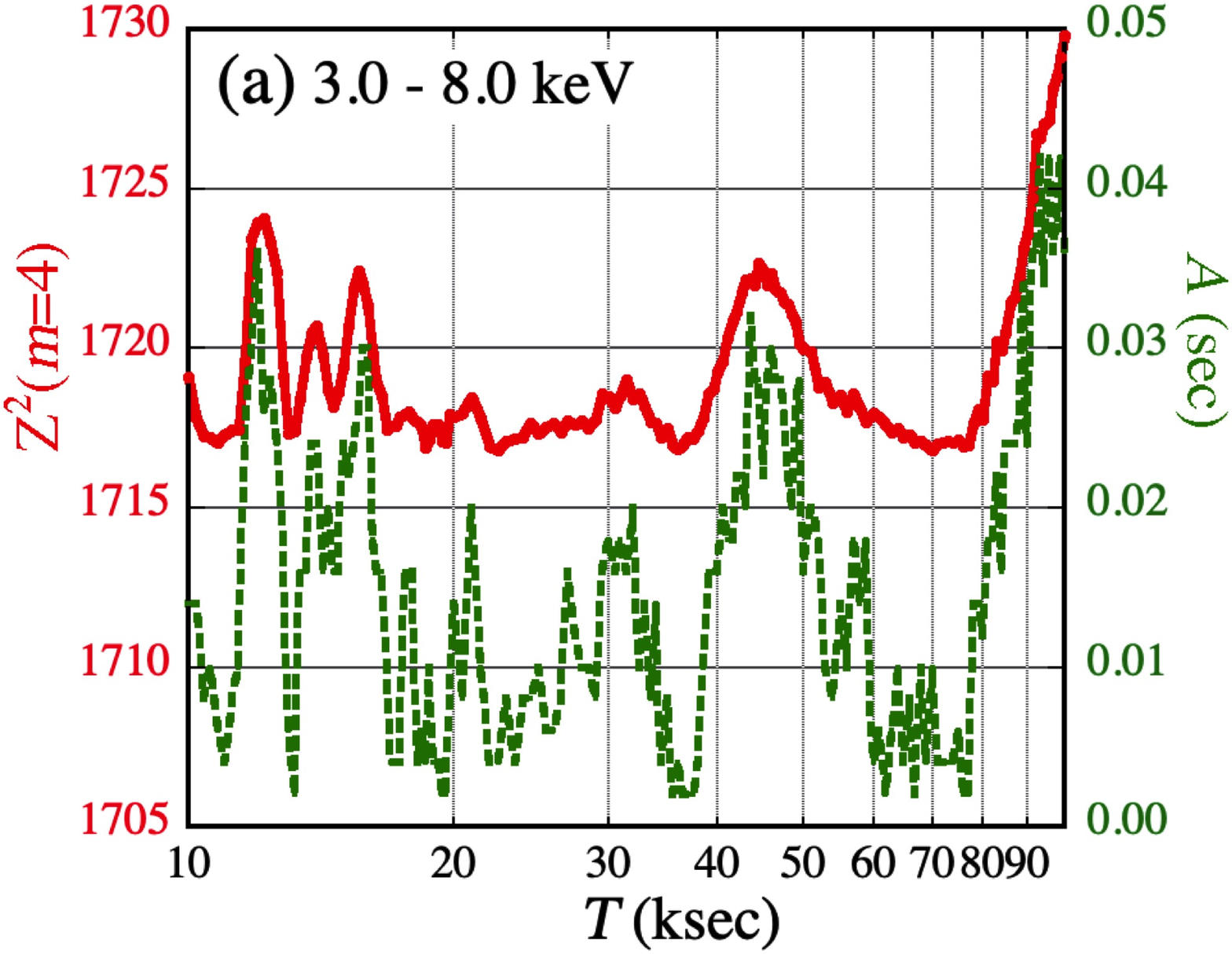}
  \includegraphics[width=60mm]{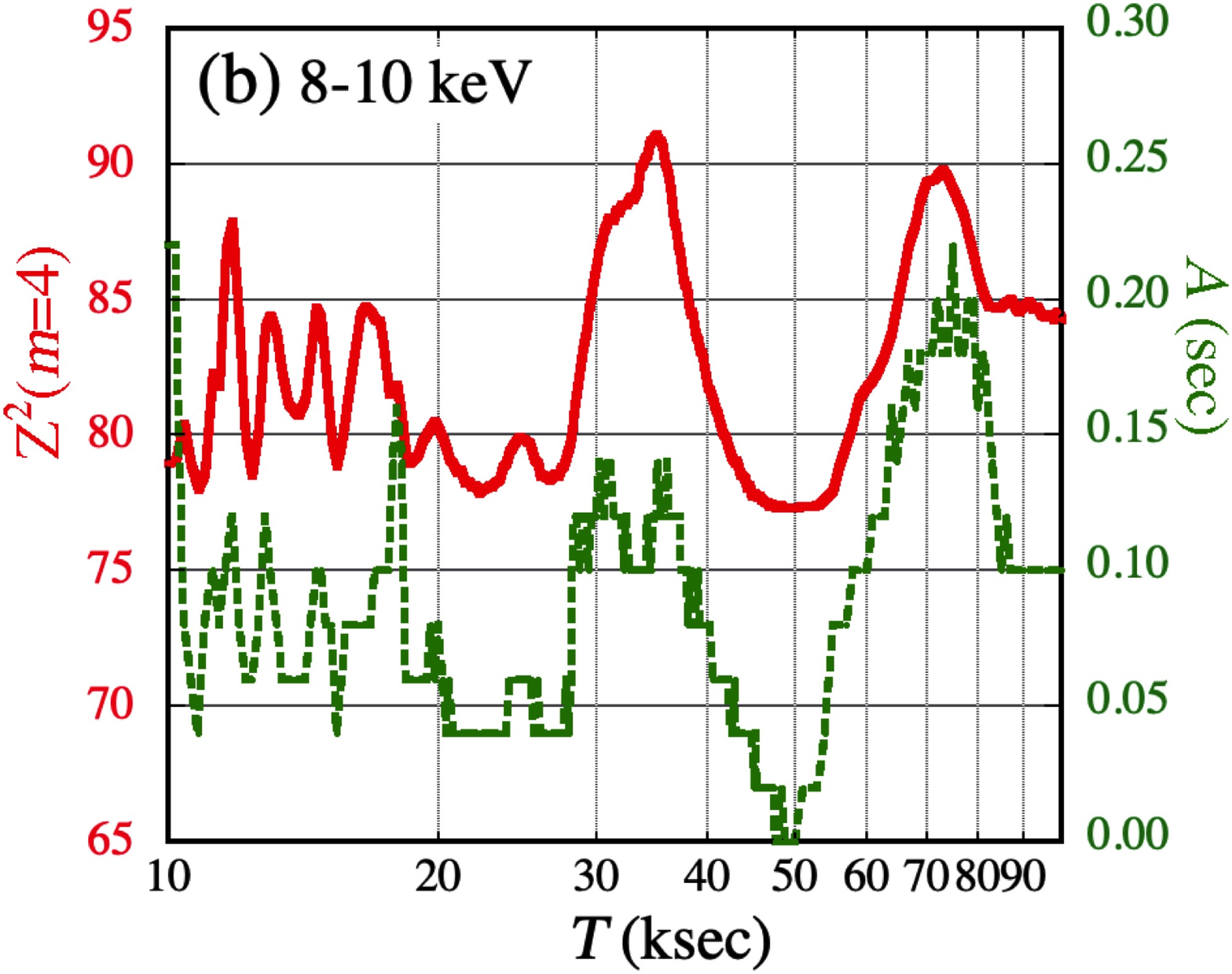}
\end{center}
\caption{The $m=4$ DeMDs in 
3--8 keV (panel a) and 8--10 keV (panel b).
The solid red curves show the maximum $Z_4^2$ values (left ordinate),
whereas the dashed green curves give $A$ (right ordinate)
 that maximizes $Z_4^2$.
 }
\label{fig9:Tscan_below_10keV}
\end{figure}

  \begin{table*}
   \begin{center}
 \caption{Basic pulse properties of \oneE\ in four  energy intervals,
 as observed with \NuS\ in 2009.}
 \renewcommand{\arraystretch}{0.95}
 \begin{tabular}{lccccccccc}
 \hline \hline 
  Energy& Stage$^{\rm a}$&  $P$   &  $Z_4^2$ & $\delta Z_4^2$ $^{\rm b}$
                                                       & $T$  & $A$ or $A_0$ &$\psi$ or $\psi_0$ &  PF $^{\rm c}$ & Reference$^{\rm d}$\\
 (keV)   &                    &  (sec)   &             & 
                                                          &  (ks) &  (sec)  & (deg)   & \\
 \hline \hline 
              & No Demod. &2.086714 &  56.18   &  ---         &  ---      & ---      & ---      & 14.2& 
              Fig.\ref{fig10:Tscan_above_10keV} black\\
8--70     & Simple Dem.& 2.086710  &  68.43   & 12.25   &  34.5  & 0.14  & 170   & 15.9 \\
               &EDPV         & 2.086708 &   108.38   & 52.20   &  36.0  & 0.51  & 190   & 20.5\\
\hline  \hline
            & No Demod.   &2.086708  &  74.20  &  ---       &  ---      & ---      & ---     & 35.4 &   
            Fig.\ref{fig4:Pr} red\\
8--10    & Simple Dem.& 2.086710  & 91.11   &  16.91   &  35.0   & 0.12   &200   & 39.6& 
Fig.\ref{fig9:Tscan_below_10keV}b, Fig.\ref{fig11:EDPV_guess}c \\
             &EDPV           &( 2.086710 & 91.81   &  17.61     &  35.0   & 0.43  & 210  & 40.0 ) $^{\rm e}$ \\
\hline  
            & No Demod. &2.086710  &  55.72     &  ---         &  ---      & ---      & ---    & 17.6 & Fig.\ref{fig3:PG} red\\
8--25    & Simple Dem.& 2.086708 & 72.95& 17.23  &  36.0  & 0.19   & 150   &20.7
& Fig.\ref{fig7:Tscan_8-25keV}, eqs.(\ref{eq:demodpara_8-25keV}),(\ref{eq:demodpara2_8-25keV})\\
             &EDPV           & 2.086708  & 92.18   &  36.46   &  36.0   & 0.52  & 190   & 23.3   & \\
\hline 
              & No Demod.    &2.086708  &  20.41  &  ---        &  ---   & ---    & ---   &9.3\\
10--40   & Simple Dem.& 2.0867010  &  33.87  &  13.46   & 38.0  & 0.19  & 310   &13.3 \\
              &EDPV             &  2.086706 &  59.70  &  39.29   & 38.0   & 0.59   &150   &18.6 \\
\hline 
             & No Demod.     &2.086710   & 24.69  &  --- &  --- & ---  & --- &14.9& Fig.\ref{fig4:Pr} blue\\
25--70  & Simple Dem.& 2.086714  & 31.10  &  6.41  &  35.0   & 0.12  & 5       & 17.2 \\
             &EDPV            & 2.086714  & 39.18  &14.49  &  38.5 & 0.56  & 160   &20.1\\
 \hline 
 \end{tabular}
 \label{tbl:Z2_summary}
 \smallskip
 \begin{itemize} 
 \setlength{\baselineskip}{4mm}
  \item[$^{\rm a}$]: ``NoDemod."= th 0th stage, without timing correction; 
  ``Simple Dem."= the 1st stage, with corrections using equation~(\ref{eq:modulation}), 
  and  the parameters in the 6th to 8th columns;
and `` EDPV"= the 2nd stage, with  corrections using equation~(\ref{eq:EDPV}) and the parameters
 in Table~\ref{tbl:EDPV_bestpara} .
 \item[$^{\rm b}$]: Increment in $Z_4^2$  from the ``NoDemod." value.
 \item[$^{\rm c}$]: The pulsed fraction in percent, 
 of which the errors are given in Fig.~\ref{fig5:LDUD}.
 \item[$^{\rm d}$]: Cross references within the paper. 
  Fig.~\ref{fig14:EDPV_Tscan},  Fig.~\ref{fig15:EDPV_PG_3cases}, 
and Fig.~\ref{fig16:EDPV_Pr_3cases} apply to all entries (except 8--10 keV) with ``EDPV".
 \item[$^{\rm e}$]: Because of the low energy range, the EDPV corrections have  rather small effects.  
  \end{itemize}
 \end{center}
  \end{table*}  

\subsection{Demodulation in various energy bands}
\label{subsec:demod_various_energies}

In \S~\ref{subsec:demod_8-25keV}, 
we reconfirmed the \Su\ result using the  \NuS\ data,
because $Z_4^2$ (and the associated PF) increased significantly 
via  demodulation with $T=36$ ks.
However, these are limited to the 8--25 keV band.
We hence attempted the demodulation analysis in other energy ranges.

\subsubsection{Results in $<10$ keV }
\label{subssubec:demod_SFC}

In the 3--8 keV  energy band where the SXC dominates,
neither Fig.~\ref{fig6:DPP} 
nor Fig.~\ref{fig8:dblfld_6bands} give evidence of 
 significant  pulse-phase fluctuations.
To be more quantitative,
we calculated the DeMD in 3--8 keV,
and show it  in Fig.~\ref{fig9:Tscan_below_10keV}a.
Thanks to the high signal-to-nose ratio attained by \NuS,
the pulsation is  strongly detected,  
with $Z_4^2=1715.7$ and the PF of 48.6\%, 
both without demodulation.
The DeMD shows several peaks,
but the highest one at $T\sim 100$ ks is  close, 
in period,  to the observation length of 151 ks,
so this would not be regarded as a periodic variation.
Furthermore, the associated value of $A = 0.05$ sec 
is only 2.5\% of one pulse cycle.
The 2nd highest one at $T=11.8$ ks is probably instrumental,
because it is just twice the orbital period of \NuS\ (5.8 ks).
The third one at $T\sim 45$ ks is eye catching,
but  it has only $\delta Z_4^2 = 6.9$,
and  $A = 0.03$ sec which is only 1.5\% of a pulse cycle.
Therefore, we regard this peak and the 4th one at $T\sim 16$ ks
as due to Poisson noise.
We do not see any $Z_4^2$ enhancement at $T \sim 36$ ks, either.
Thus, the soft X-ray pulses below 8 keV, dominated by the SCX,
are  free from  pulse-phase modulations,
at 36 ks or any other period studied here.
We quote a typical upper limit of $A \lesssim 0.04$ sec, 
for the 36 ks pulse-phase modulation in 3--8 keV.

\begin{figure*}
\centerline{
\includegraphics[width=53mm]{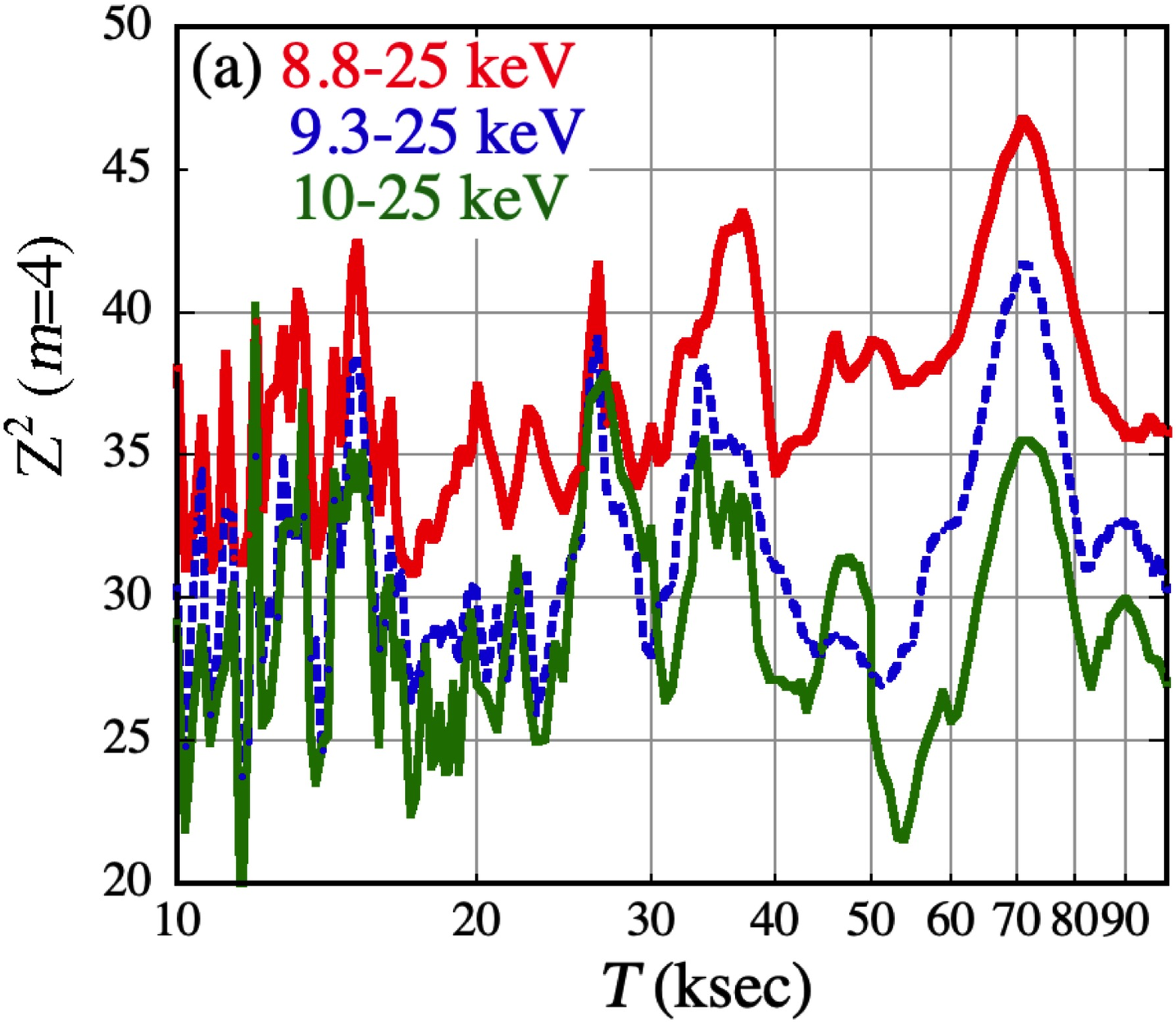}
\includegraphics[width=55mm]{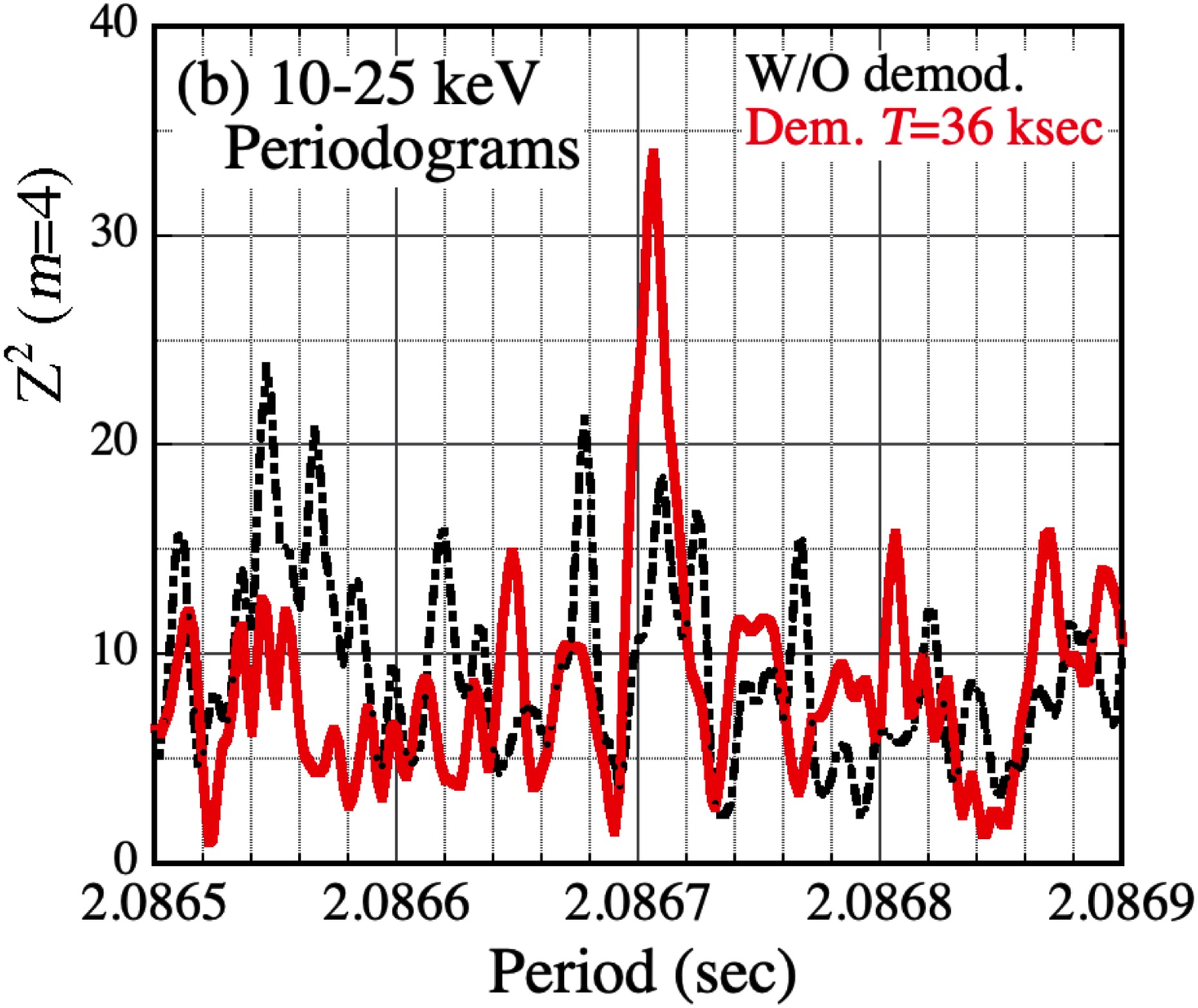}
\includegraphics[width=55mm]{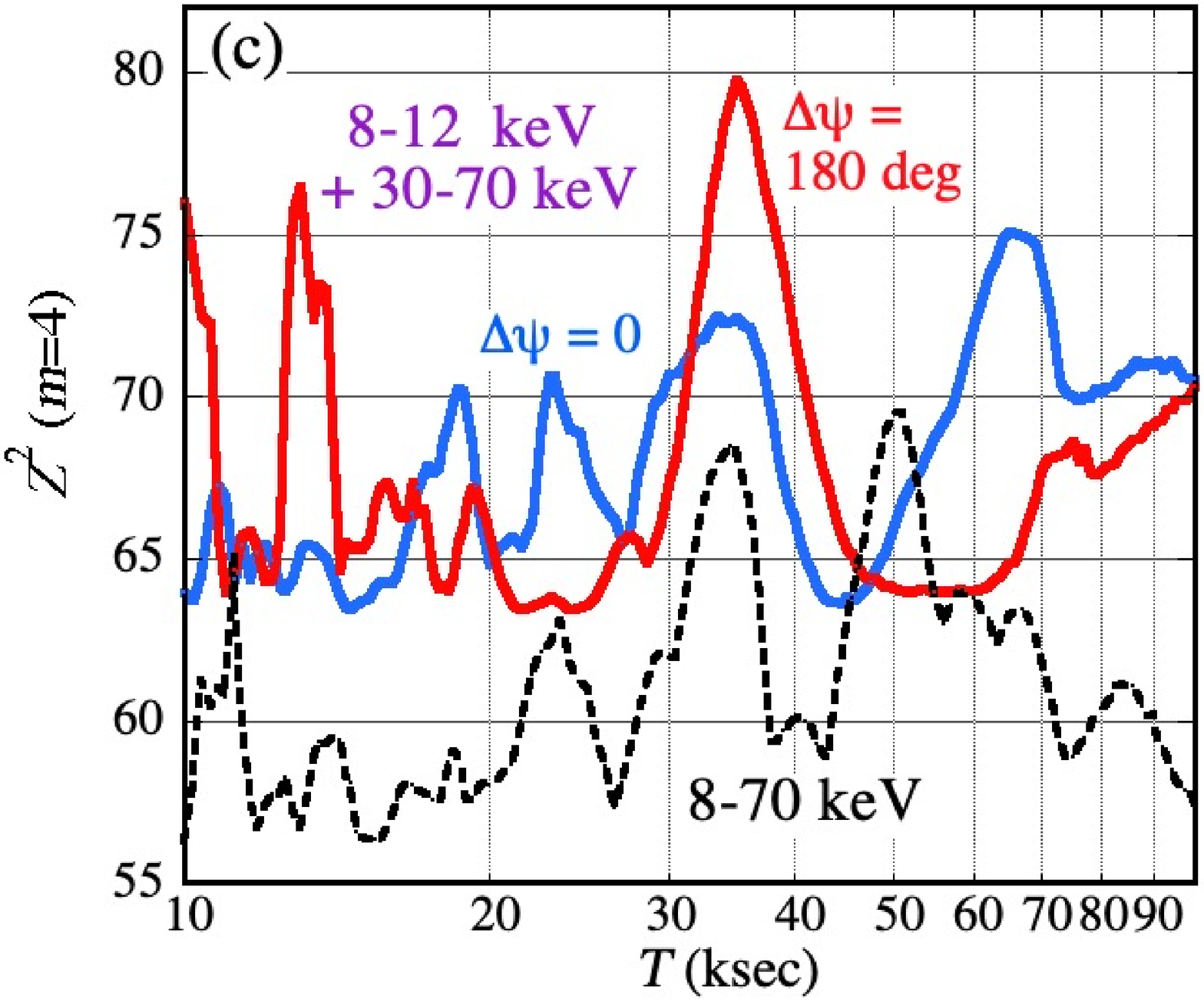}
}
\caption{
(a) The DeMDs  from  three energy intervals above 10 keV. From top to bottom, 
8.8--25 keV (solid red), 9.3--25 keV (dashed blue), and 10--25 keV (solid green).
(b) The 10--25 keV $m=4$ pulse periodograms, 
shown on the same period scale as Fig.~\ref{fig3:PG}. 
The dashed black trace is before the demodulation.
The solid red trace is after the demodulation,
using $A=0.5$ sec, $\psi=150^\circ$, and $T=36.0$ ks.
(c) The dashed black trace shows the 8--70 keV DeMD,
whereas the blue one uses two disjoint energy intervals,
8--12 keV plus 30--70 keV. 
The red one is the same, 
but further incorporating a modulation-phase jump
by $\Delta \psi =180^\circ$ between 12 keV and 30 keV.
 }
\label{fig10:Tscan_above_10keV}
\end{figure*}

Figure~\ref{fig9:Tscan_below_10keV}b shows the $m=4$ DeMD,
together with the behaviour of $A$, 
in  8--10 keV  where the two spectral components have 
comparable contributions (Fig.~\ref{fig1:spec}).
The $T=36$ ks feature has  emerged clearly,
of which the parameters are summarised in Table~\ref{tbl:Z2_summary}.
Compared to equation~(\ref{eq:demodpara_8-25keV})
describing the 8--25 keV results, 
$T=35.0^{+2.1}_{-4.8}$ ks is consistent,
$A=0.12 \pm 0.06$ sec is somewhat smaller, 
and $\psi=200^\circ \pm 30^\circ$ could be larger. 
The 72 ks hump is also seen.
The 36 ks phase modulation in this energy band
is visualised by a double-folded map  in Fig.~\ref{fig8:dblfld_6bands}c,
which is similar to that in 8--25 keV.
The pulse behaviour thus changes  at  $\sim 8$ keV,
rather than at  $\sim 12$ keV where the HXC and SXC cross over.

\subsubsection{Results in $>10$ keV}
\label{subsubsec:demod_HXC}

We next examine higher-energy data 
for the pulsation and its 36 ks phase modulation,
trying to solve the puzzle of the reduced PF.
With the UD kept at 25 keV as in Fig.~\ref{fig7:Tscan_8-25keV},
we hence raised  the LD, from 8 keV to 10 keV stepwise.
Figure~\ref{fig10:Tscan_above_10keV}a 
shows how the DeMD changed  in this course.
When the LD is raised, {\it e.g.},  to 8.8 keV,
the 36 ks feature became somewhat weaker,
whereas the 72 ks  hump more prominent.
As the LD is further raised to  9.3 keV,
the 70 ks hump still remained,
but the 36 ks peak nearly disappeared.
In addition, the $Z_4^2$ values on average decreased
more than is expected from  the count-rate decrease
(by 24\% from that in 8--25 keV) via equation~(\ref{eq:PF_vs_Z2});
this implies a decrease in the PF.
These trends further continued to 10--25 keV,
where the 36 ks peak became no longer visible.

Thus, some drastic changes in the pulsation is
likely to take place as the LD is raised from 8 keV to 10 keV,
in such a way that the correction with  equation~(\ref{eq:modulation}),
which worked fine for  energies 8--25 keV,
becomes less effective.
Nevertheless, an  alternative possibility,
that the  PF in $>10$ keV is intrinsically low, is unlikely, 
because we  keep observing the 72 ks hump,
which may be related to the 36 ks feature.
Moreover,  when $T=36.0$ ks and $P=P_0$ are fixed,
the 9.3--25 keV and 10-25 keV DeMDs 
yield $\delta Z_4^2=13.84$ and  $\delta Z_4^2=12.63$, respectively,
which are not too small.
This becomes clear in Fig.~\ref{fig10:Tscan_above_10keV}b,
which compares the 10--25 keV periodograms
before and after the demodulation.
Thus, the  correction with equation~(\ref{eq:modulation})
does enhance the pulsation, 
but some other values of $T$ give still  larger $Z_4^2(P_0)$
as in  Fig.~\ref{fig10:Tscan_above_10keV}a (green).

As we suspected in \S~\ref{subsub:profiles}, these results suggest
that the intrinsic PF in the intermediate energy ranges would be
in reality  much higher than is implied by  simple epoch-folding results,
and the pulse coherence is degraded by some decoherence processes
that cannot be fully described  by  equation~(\ref{eq:modulation}).
The most likely possibility,
as  suggested by Fig.~\ref{fig5:LDUD},
is that the pulse behaviour including its 36 ks phase modulation
is  considerably energy dependent,
so the  pulses in different energies partially cancel out
when the data are accumulated over a wide energy.

The above inference has been reinforced by 
Fig.~\ref{fig10:Tscan_above_10keV}c.
The dashed black trace represents the 8--70 keV DeMD,
where the baseline is relatively high at $Z_4^2 \sim 56$
but the 36 ks peak ($Z_4^2 \sim 68$) is not very strong.
We then found that the pulse significance generally increases
when we discard some intermediate energy intervals.
As indicated by a blue curve,
this effect  became most prominent 
when the 12--30 keV interval is discarded 
and the remaining two energy ranges,  
8--12 plus 30--70 keV, are used jointly.
In spite of a count rate decrease by 40\%, 
on average  $Z_4^2$ became higher,
and the 36 ks peak increased to $Z_4^2 =72.42$.
The demodulated PF increased
from 15.9\% to $21.5\pm 2.9 \%$.
Therefore, the suspected decoherence effects are
thought to be  significant in the excluded 12--30 keV energy range.
A surprise was the red DeMD,
where the 36 ks peak drastically increased to $Z_4^2=79.78$.
This curve was obtained in the same manner as the blue one,
but assuming that  $\psi$ in equation~(\ref{eq:modulation})
changes by $\Delta \psi = 180^\circ$ between 12 keV and 30 keV.
When $\Delta \psi$ is allowed to vary, $Z_4^2$ became 
maximum for $\Delta \psi = 170^\circ \pm 50^\circ $.
Thus, the decoherence effects are inferred
to involve a {\it phase reversal} in $\psi$,
across the intermediate energy range.

We are here brought back to the issue that has so far been left behind;
the energy dependent variations
in the pulse phase. 
To be  precise,  we need to consider two independent factors.
One is the simple energy dependence of the pulse phase,
as indicated by Fig.~5 of CZ20 and our Fig.~\ref{fig4:Pr}.
The other is the  more complex possibility 
revealed by Fig.~\ref{fig10:Tscan_above_10keV}c,
that $\psi$ in equation~(\ref{eq:modulation}),
and possibly $A$ as well, depend on energy.
We call these two mechanisms collectively 
{\it energy-dependent pulse variation (EDPV)} effects.

\subsection{Energy dependent pulse variation (EDPV)}
\label{subsec:demod_Lorentz}

Assuming that $T$ is energy independent,
the  idea of EDPV can be formulated
by modifying equation~(\ref{eq:modulation}) as
\begin{equation}
\delta t =  P \cdot S(E)+ \tilde A(E) \sin \left[ 2\pi t/T - \tilde{\psi}(E) \right] ~
\label{eq:EDPV}
\end{equation}
where $E$ is the photon energy in keV,
and $S(E)$ ($-1 \leq S<1$, most likely $0\leq S\lesssim 0.25$) 
describes the energy-dependent but $T$-unrelated pulse-phase shifts
in units of the pulse cycle.
The variations depending on both $E$ and the modulation phase $\Psi$ 
are taken into account by $\tilde A(E)$ and $\tilde{\psi}(E)$,
which generalise $A$ and $\psi$ in equation~(\ref{eq:modulation}), respectively.
This $\tilde{\psi}(E)$ includes the modulation-phase change
by $\Delta \psi$ indicated by Fig.~\ref{fig10:Tscan_above_10keV}c.
In a double-folded map, 
$S(E)$ and  $\tilde{\psi}(E)$  respectively represent
energy-dependent horizontal and vertical displacements of the pulse pattern;
the latter has no meaning when $A= 0$.
If we regard Fig.~\ref{fig3:PG} and Fig.~\ref{fig4:Pr}
as the 0th stage of the pulsation study,
the demodulation analysis using equation~(\ref{eq:modulation})
can be considered as the 1st stage,
and the introduction of equation~(\ref{eq:EDPV}) means
that we proceed to the 2nd stage.

\begin{figure*}
\centerline{
\includegraphics[width=71mm]{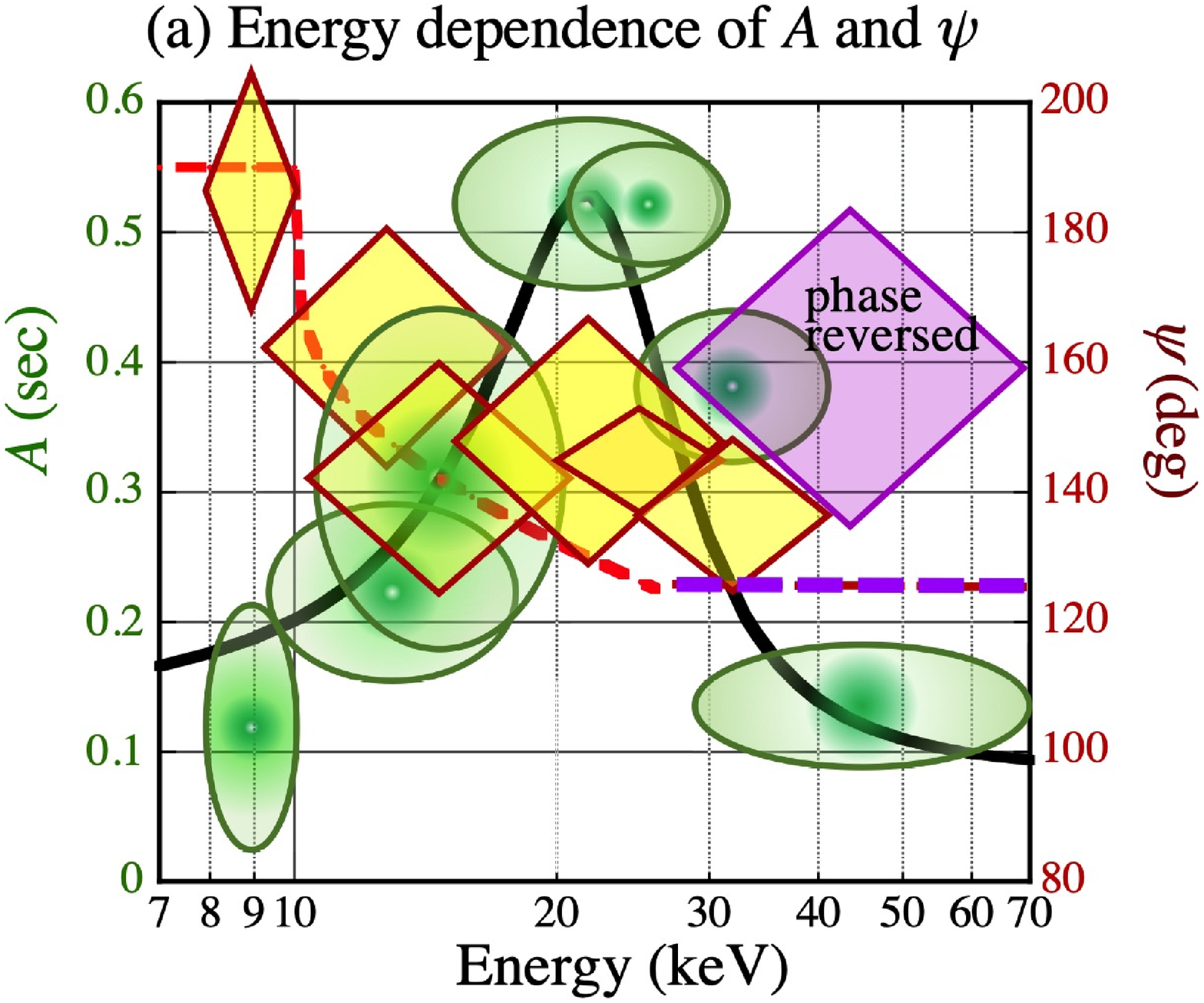}
\includegraphics[width=58mm]{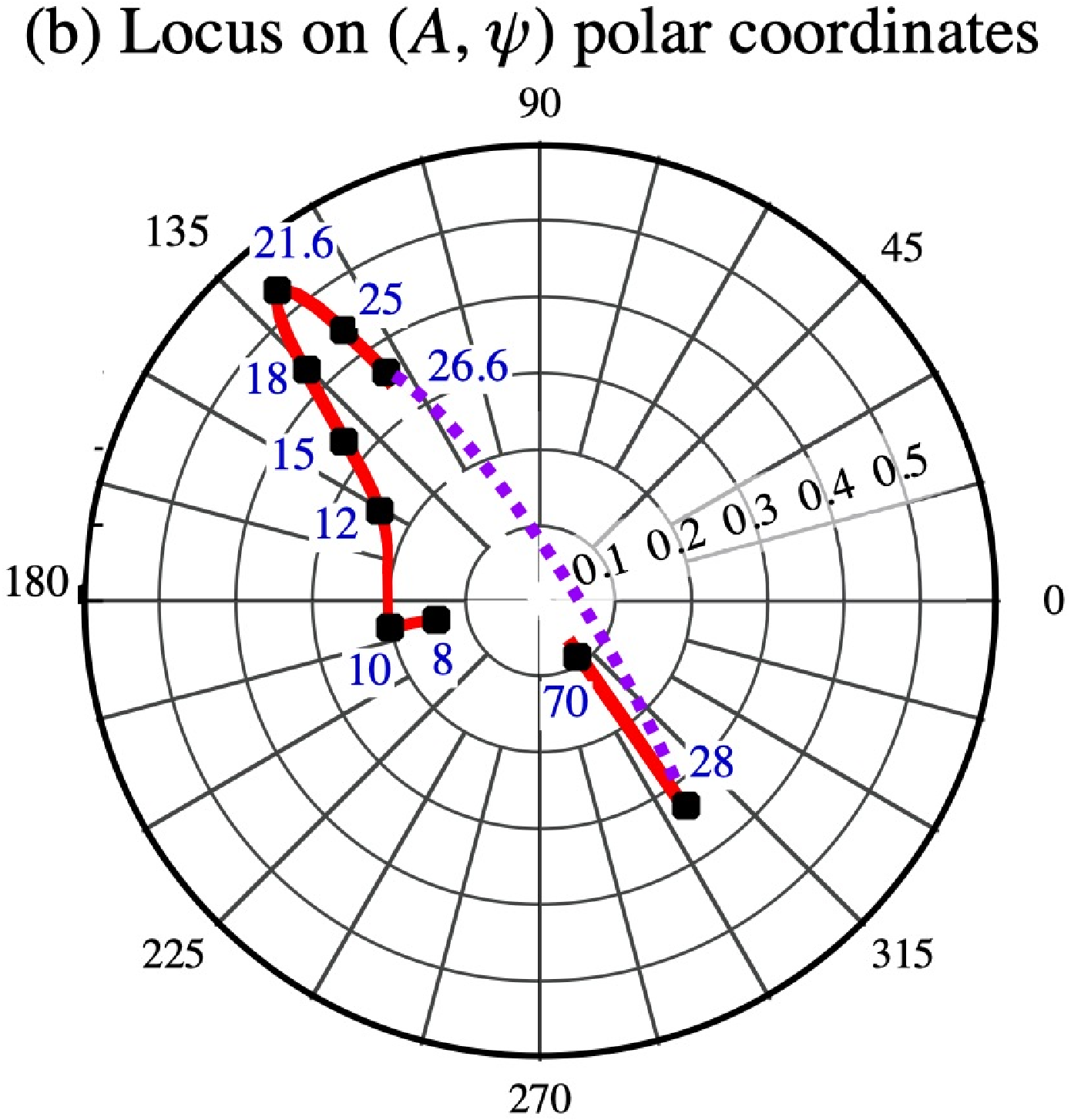}
}
\centerline{
\includegraphics[width=57mm]{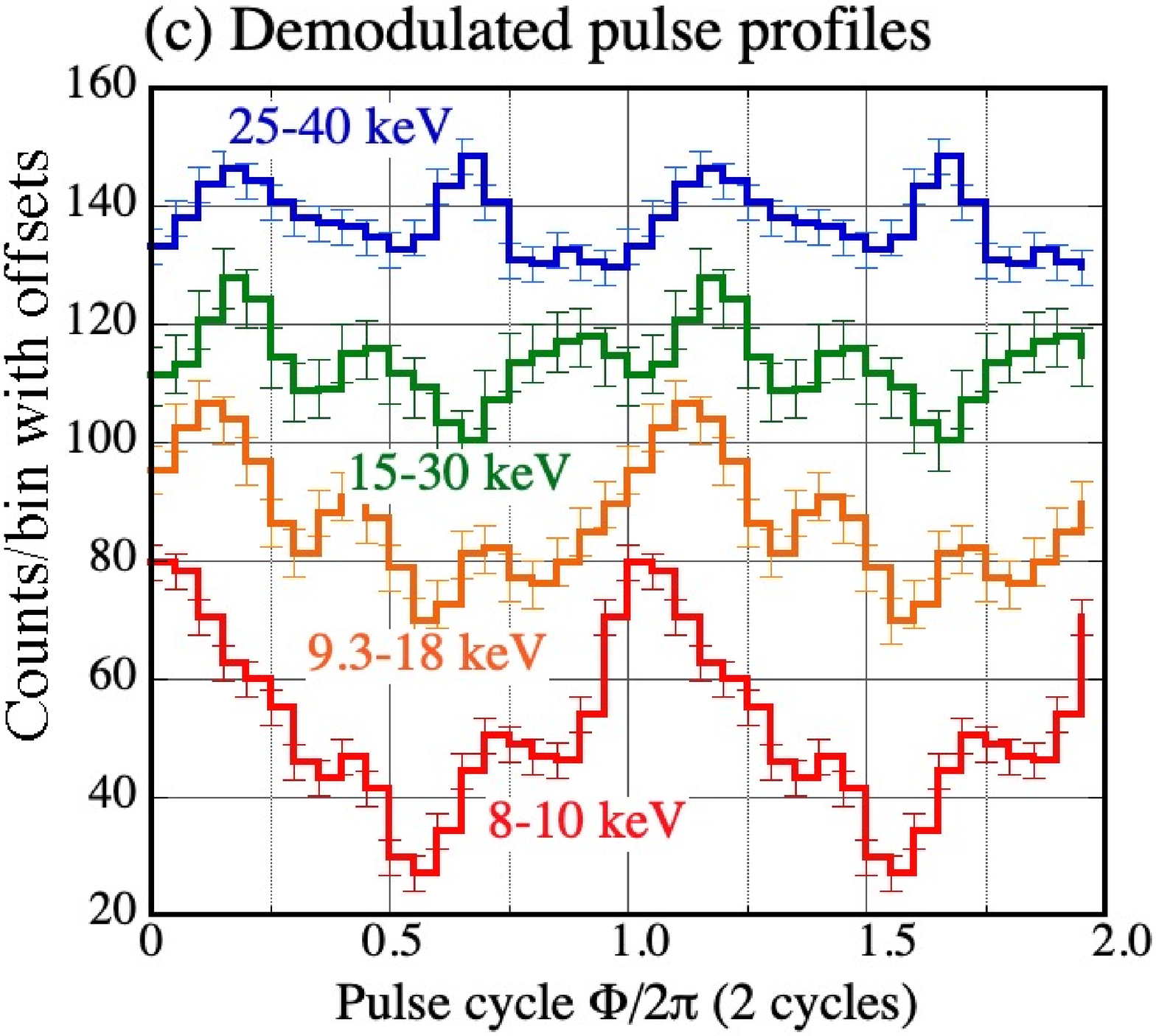}
\includegraphics[width=26mm]{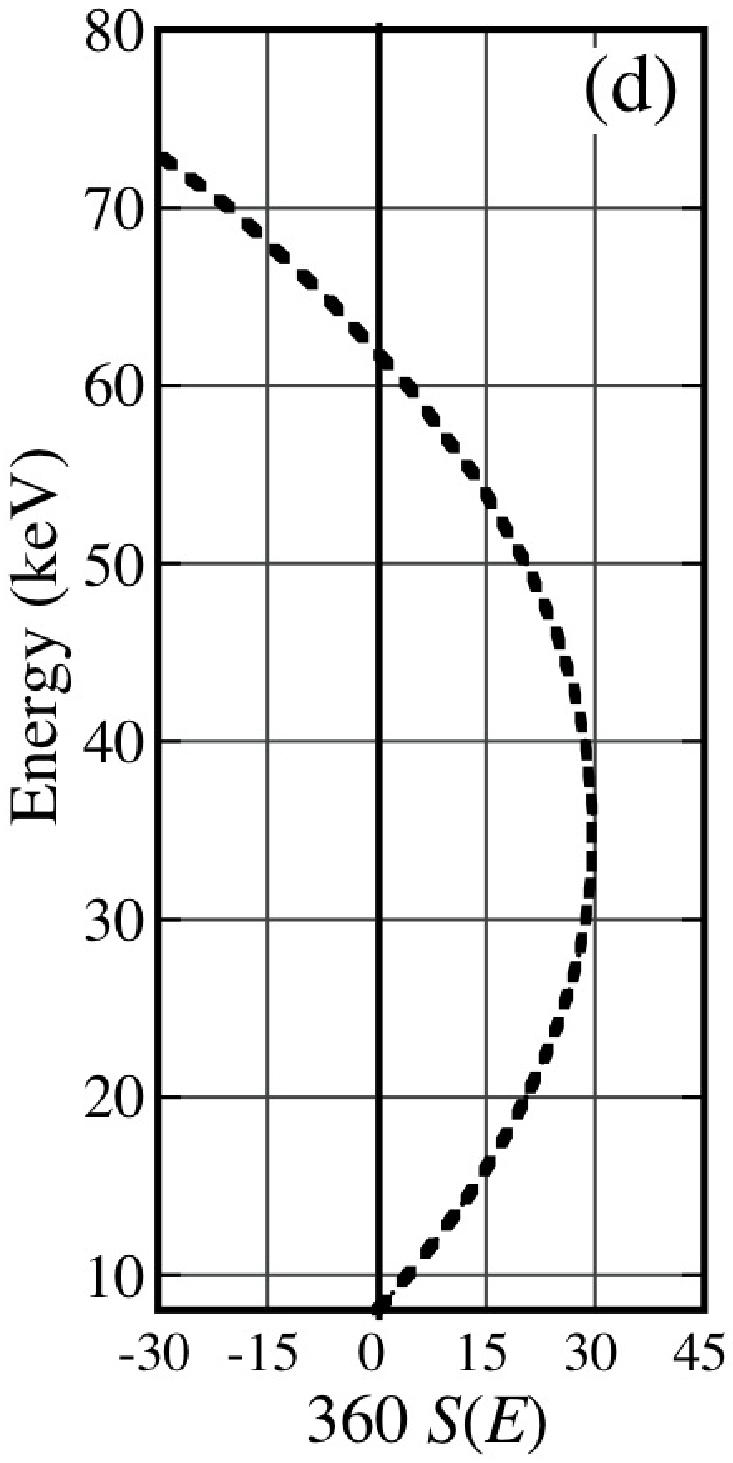}
}
\caption
{ 
(a) The values of $A$ (green ellipses) and $\psi$ (yellow diamonds)
that maximize $Z_4^2$ at $T=36.0$ ks (fixed) and $P \sim P_0$, 
calculated in various energy bands.
The vertical heights of these symbols represent  $\pm 1 \sigma$ errors,
and their lateral widths the  employed energy ranges.
The 27--70 keV $\tilde{\psi}$ data point (purple diamond) is shown by adding $180^\circ$.
The thick black curve and  the dashed red line describe
$\tilde{A}(E)$ and $\tilde{\psi}(E)$, respectively,
using the optimum parameters in Table~\ref{tbl:EDPV_bestpara}.
At $\geq$ 26.6 keV, $\tilde{\psi}(E)$ makes a $180^\circ$ phase flip
as indicated by the dashed purple line.
(b) Locus of the best-estimated $\tilde{A}(E)$ and $\tilde{\psi}(E)$,
shown on the $(A, \psi)$ polar coordinates.
The numbers in blue are the energy in keV.
The  dotted purple line indicates a possible trajectory across the $\psi$ reversal. 
(c) Pulse profiles in several energy bands used in panel (a),
shown again with the running average, 
and vertical offsets to avoid overlap.
They were obtained through the demodulation using $T=36.0$ ks,
and the optimum $A$ and $\psi$ determined in respective bands.
(d) The form of $S(E)$ in equation~(\ref{eq:S(E)}),
calculated using the parameters in Table~\ref{tbl:EDPV_bestpara}
and shown in unit of degree, after multiplication by $360^\circ$.
The energy $E$ is taken as the ordinate.
} 
\label{fig11:EDPV_guess}
\end{figure*}

 \vspace*{-1mm}
\subsubsection{Formalism}
\label{subsubsec:EDPV_formalism}
To implement equation~(\ref{eq:EDPV}) into our analysis,
we must come up with some appropriate forms of
$\tilde A(E)$, $\tilde{\psi}(E)$, and $S(E)$,
based on the data themselves,
as we can invoke  neither theoretical predictions,
nor past observation of a similar kind.
Figure~\ref{fig10:Tscan_above_10keV} is important,
but not yet informative enough.
We hence repeated the demodulation analysis in various energy ranges 
between 8  and 70 keV,
chosen to be relatively narrow as long as
$Z_4^2(P_0)$ for $T=36.0$ ks exceeds 25.0.
Figure~\ref{fig11:EDPV_guess}a 
summarises $A$ and $\psi$
that  maximize $Z_4^2(P_0)$ at  $T=36.0$ ks.
It thus reveals strong  energy dependences
in both  $A$ and $\psi$;
$A$  increases from $\sim 0.1$  at $\sim 10$ keV,
to $\sim 0.5$  at $E \sim 20$ keV,
and then returns to $\sim 0.1$ at $E>30$ keV.
These variations in $A$ naturally explain 
why the pulse significance increased in Fig.~\ref{fig10:Tscan_above_10keV}c
by excluding the 12--30 keV interval.

As indicated by yellow diamonds in Fig.~\ref{fig11:EDPV_guess}a,
$\psi$ starts from $\sim 180^\circ$,
and gradually decreases towards higher energies up to $\sim 30$ keV,
but the drop meantime is only $\sim 60^\circ$.
In particular, the change from 12 keV to 30 keV 
is only $\sim 20^\circ$,
much smaller than  $\Delta \psi \sim 180^\circ$ 
suggested by Fig.~\ref{fig10:Tscan_above_10keV}c.
Instead, as shown by a purple diamond,
the 27--70 keV data point implies a sudden change in $\psi$:
it was in reality obtained at $\psi=340^\circ \pm 25^\circ$,
but for the presentation,
it is shown after shifting by $180^\circ$ along the ordinate.
Since this result is likely to reflect the reversal in $\tilde{\psi}$,
we returned to Fig.~\ref{fig10:Tscan_above_10keV}c
 and changed the upper bound $E_3$ 
of the excluded energy region.
As $E_3$ was gradually lowered from 30 keV,
the optimum $\Delta \psi$ remained at $150^\circ-180^\circ$,
until $E_3 \lesssim 25$ keV
when the advantage of  assuming $\Delta \psi \ne 0$ quickly diminished.
Considering these,
we may regard $\tilde{\psi}(E)$ as 
gradually decreasing from $E \sim 10$ keV to $\sim 25$ keV,
followed by a $\sim 180^\circ$ jump (phase reversal)
 in the narrow  25--30 keV interval.

{Just for a cross confirmation,
panels (d)-- (f) of Fig.~\ref{fig8:dblfld_6bands}
show double-folded maps in three energy bands above 10 keV.
Like in panels (b) and (c),
the 10--12 keV map in (d) again shows a wiggling ridge,
starting from $(\Phi/2\pi, \Psi/2\pi) \sim (1.1, 0.2)$ to run down to $(1.3, 0.9)$.
A difference from the 8--10 keV  map
is a  feature at $(\Phi/2\pi, \Psi/2\pi) \sim (1.4, 0.2)$,
which may have disturbed the simple demodulation.
Panel (e) in 12--25 keV is more complex and ambiguous,
but one of possible interpretations could be
that the pulse ridge leads from $(0.9, 0.2)$  to $(1.2, 0.9)$,
with a larger lateral swing as suggested by Fig.~\ref{fig11:EDPV_guess}a, 
and another wiggling feature runs roughly in parallel,
from $(1.3, 0.1)$ to $(1.6,0.7)$.
In the 25--70 keV map, which is also rather noisy,
the main ridge may have returned more straight at 
$\Phi/2\pi \sim 1.2$, which is  somewhat delayed,
as represented by $S(E)$, from those in 3--8 keV.
Furthermore, the ridge could be most advanced at $\Psi/2\pi \sim 0.6$,
which is opposite to the behaviour in  panels (b) and (c).
This is consistent with the phase reversal in $\tilde{\psi}(E)$ at $\sim 25$ keV.
Thus, the double-folded maps, though not quantitative,
appear generally consistent  with Fig.~\ref{fig11:EDPV_guess}.

From these preparations,
$\tilde A(E)$ may be regarded as 
consisting of a constant floor at $A \sim 0.1$,
plus a sharp enhancement centred at $\sim 20$ keV
which has  a height of  $A \sim 0.5$
and a width of $\sim 10$ keV.
As a  simple analytic form to empirically describe such a profile,
we choose a Lorentzian plus a constant, 
and describe as
\begin{equation}
\tilde A'(E) = A_0 \left[  a_{\rm f} + 
\frac{1- a_{\rm f}}{   1.0+\left\{ (E-E_{\rm c})/E_{\rm w} \right\}^2  } \right]~.
\label{eq:A(E)}
\end{equation}
using three parameters;
the centroid $E_{\rm c}$ and the width $E_{\rm c}$ of the Lorentzian,
and a dimensionless amplitude floor $a_{\rm f}$.
In addition,  $A_0$ replaces $A$ in equation~(\ref{eq:modulation}).
As  superposed in Fig.~\ref{fig11:EDPV_guess}a on the data,  
$\tilde A(E)$ reaches the maximum of $A_0$ at $E=E_{\rm c}$,
and decreases at $E> E_{\rm c}$  to approach $A_0a_{\rm f}$.

Based on the considerations performed so far,
$\tilde \psi(E)$  is modeled as as 
\begin{equation}
\tilde \psi(E) = 
\left\{
\begin{array}{ll}
 \psi_0                                                    & (E \leq 10)\\
\psi_0 - \psi_{\rm d} \left[ (E-10) /(E_{\rm d}-10)\right]^\gamma & (10 < E < E_{\rm d} )\\
\psi_{\rm 0}- (\psi_{\rm d} +\Delta \psi )                 & (E_{\rm d} <E)
\end{array}
\right.
\label{eq:psi(E)}
\end{equation}
using five parameters, $ \psi_0$,  $\psi_{\rm d}$,  $ E_{\rm d}$, 
$\gamma$,  and $\Delta \psi$.
Here, $ \psi_0$ replaces $\psi$ in equation~(\ref{eq:modulation}),
and specifies the initial modulation phase at $E \leq 10$ keV.
From $E=10$ keV to $E=E_{\rm d} \sim 25$ keV,
$\tilde{\psi}$ is assumed to decrease by $\psi_{\rm d}$,
as a power-law function of $E$ with an index $\gamma>0$.
At $E=E_{\rm d}$,
$\tilde{\psi}$ reaches $\psi_{\rm 0}- \psi_{\rm d}$ 
where it makes the phase jump by $\Delta \psi \sim 180^\circ$,
and stays there afterwards.
Assuming no energy dependence after the phase jump 
is justified by the constancy of $\Delta \psi$
when we scanned $E_3$ from 30 keV to 25 keV.
The functional form of  $\tilde \psi$ is also superposed
in Fig.~\ref{fig11:EDPV_guess}a on the data,
where the purple portion is subject to the phase reversal.

To get an idea on $S(E)$,
we show in Fig.~\ref{fig11:EDPV_guess}c the pulse profiles folded at $P_0$,
in four representative (and slightly overlapping) energy bands
chosen from those used in panel (a).
Unlike Fig.~\ref{fig4:Pr},
these profiles have been derived through demodulation,
using equation~(\ref{eq:modulation}) assuming $T=36.0$ ks, 
together with the optimizing $A$ and $\psi$ in each energy band.
Here, the simple demodulation has already 
made the pulse profiles  behave much more systematically 
as a function of energy than before.
Among the lower 3 bands, 
the main pulse peak and the pulse minimum  
are both  seen to shift positively as a function of energy,
whereas the dependence appears to saturate
(except the sub peak in 25--40 keV emerging at $\Phi/2\pi \sim 0.6$)
when the highest two bands are compared.
Therefore, we have chosen to express $S(E)$ 
using a simple parabolic function as
\begin{equation}
S(E) = 
\frac{R/360  }{ (E_{\rm piv}-8)} (E-8)(E_{\rm piv}-E)
\label{eq:S(E)}
\end{equation}
for $E>8$ keV and $S(E)=0$ for $E<8$ keV,
with two free parameters;
the pivot energy $E_{\rm piv}$ where the modulation phase returns to that at 8 keV,
and the rate of the pulse-phase change at 8.0 keV,
$R \equiv \left\{ d S(E)/dE \right\}_{\rm 8 keV}$ in units of degree per keV.
The pulse-phase lag increases for $8 < E <(8+E_{\rm piv})/2$,
decreases for $(8+E_{\rm piv})/2 <E < E_{\rm piv}$,
and turns negative for $E > E_{\rm piv}$.
This function is illustrated in Fig.~\ref{fig11:EDPV_guess}d,
where we rotated the coordinate direction so as to simulate panel (c).
The result does not change very much,
if the start point is set at 10 keV instead of 8 keV.

\subsubsection{Optimisation of the new parameters}
\label{subsubsec:EDPV_optimisation}

The above  2nd-stage recipe with equations~(\ref{eq:EDPV}) through (\ref{eq:S(E)}) 
altogether involves 9 new parameters, 
$E_{\rm c}$, $E_{\rm w}$, $a_{\rm f}$,
$\psi_{\rm d}$, $E_{\rm d}$, $\gamma$, $\Delta \psi$,
 $R$, and $E_{\rm piv}$,
in addition to $A_{\rm 0}$ and $\psi_0$.
The 9 new parameters, once appropriately optimized, are expected 
to significantly enhance the $T=36$ ks peak in the DeMDs at any energy, 
using a common pair of $(A_0, \psi_0)$.
To achieve this goal,
we have selected an intermediate energy band of 10--40 keV,
because $\tilde A(E)$, $\tilde \psi(E)$, and $S(E)$ are
all inferred to be most  energy dependent
around these energies,
and the photon statistics are still sufficient.
Some pilot studies indicated that the data still prefer 
$\Delta \psi \sim180^\circ$ within a tolerance of  $\pm 30^\circ$,
and a rather small index as $\gamma=0.2-0.3$.
We hereafter assume $\Delta \psi =180^\circ$
and $\gamma=0.25$ both fixed,
and try to optimize the remaining 7 parameters.

We have hence trimmed $E_{\rm c}$, $E_{\rm w}$, $a_{\rm f}$, 
$\psi_{\rm d}$, $E_{\rm d}$,  $R$, and $E_{\rm piv}$, 
in addition to $A_0$ and $\psi_0$, 
so as to maximize $Z_4^2$ in the 10--40 keV interval
for $T\sim 36.0$ ks and $P\sim P_0$.
Then, their optimum values and typical errors have been determined as
 in Table~\ref{tbl:EDPV_bestpara}.
The functional forms of $\tilde{A}(E)$ and $\tilde{\psi}(E)$,
plotted in Fig.~\ref{fig11:EDPV_guess}a,
actually employ these optimum parameters.
They are seen to approximately reproduce the data behaviour,
although the best-estimated $\tilde \psi (E)$
decreases with energy somewhat less steeply than the yellow diamonds.
Also, the $S(E)$ curve in Fig.~\ref{fig11:EDPV_guess}d is drawn
using these optimum values of $R$ and $E_{\rm piv}$.
Thus, the HXC pulse phase is implied to vary with energy
by $\sim 0.08$ pulse cycles (CZ20),
independently of the 36 ks modulation phase.

Figure~\ref{fig11:EDPV_guess}b shows the locus of 
the best-estimated $\tilde{A}(E)$ and $\tilde{\psi}(E)$,
on the $(A,\psi)$ polar coordinate.
As indicated with the dashed purple line,
the phase reversal is understood as a quick motion
of the data point through $A\sim 0$,
even though the exact locus of the phase reversal is unconstrained.

\subsubsection{Results}
\label{subsubsec:EDPV_results}
Figure~\ref{fig12:EDPV_bestfit_curves} depicts 
the pulse-peak behaviour  at four representative energies,
as specified by the optimum EDPV parameters 
(Table~\ref{tbl:EDPV_bestpara}).
The plot utilises the same coordinates as  Fig.~\ref{fig8:dblfld_6bands},
where energy-dependent vertical and horizontal shifts of the pulse-peak loci
are represented by $\tilde \psi (E) $ and $S(E)$, respectively.
The straight black lines represent the pulses at $E<8$ keV,
as in Fig.~\ref{fig8:dblfld_6bands}a.
The red curve for  $E=10$ keV is similar to
the wiggling pulse-peak locus in Fig.~\ref{fig8:dblfld_6bands}c,
and the green curve for $E=22$ keV $\approx E_{\rm c}$
is seen to crudely emulate 
our particular interpretation of Fig.~\ref{fig8:dblfld_6bands}e
(although the  secondary feature is not reproduced here).
Reflecting the decrease in $\tilde{\psi}$ from 10 to 22 keV
by about $60^\circ$ (Fig.~\ref{fig11:EDPV_guess}a),
the green curve is shifted upwards  by $\sim 0.2$ cycles
compared to the red one.
The  blue curve,
to be compared with Fig.~\ref{fig8:dblfld_6bands}f,
has the opposite modulation phase compared to the green curve,
because of the phase reversal at $E \sim 27$ keV.

Below, we adopt the formalism using 
equations~(\ref{eq:EDPV})  through  (\ref{eq:S(E)}),
and the parameters in Table~\ref{tbl:EDPV_bestpara},
as the final solution of our 2nd-stage study.

\begin{figure}
\centerline{\includegraphics[width=53mm]{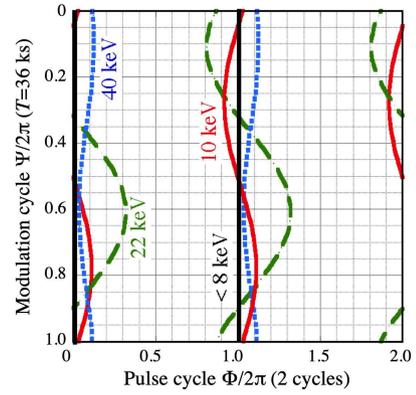}}
\caption{
Loci of the pulse peaks indicated by 
equations (\ref{eq:EDPV}) --  (\ref{eq:S(E)})
and the parameters in Table~\ref{tbl:EDPV_bestpara},
shown on the same  two-dimensional plane as Fig.~\ref{fig8:dblfld_6bands}.
Black, red, dashed green, and dotted blue are the predictions for
$<8$ keV, 10 keV, 22 keV, and 40 keV,  respectively,
to compare with panels (a), (c), (e), and (f) of Fig.~\ref{fig8:dblfld_6bands}.
 }
\label{fig12:EDPV_bestfit_curves}
 \end{figure}

  \begin{table}
   \begin{center}
 \caption{The EDPV parameters optimized in 10--40 keV.}
 \begin{tabular}{llcccc}
 \hline \hline 
 \multicolumn{2}{l}{Parameter}& \multicolumn{2}{c}{Eq. (\ref{eq:EDPV})(\ref{eq:A(E)})} & Eq. (\ref{eq:A'(E)}) & Eq. (\ref{eq:EDPV_square})$^{\rm c}$\\
    \cmidrule{3-4}
     &                       & value &  error$^{\rm a}$ & value & value\\
\hline  
 \multicolumn{2}{l}{$\tilde{A}(E)$}\\
    &$E_{\rm c}$  (keV)  & 21.6  & 0.4 & 20.5 & 21.1  \\
    &$E_{\rm w}$ (keV)   & 6.9   & 0.3  &  8.5 & 3.6 \\
    &$a_{\rm f} $            &0.16  & 0.03 &0.16 & 0.18\\
\hline  
  \multicolumn{2}{l}{$\tilde{\psi}(E)$}\\
    &$ \psi_{\rm d}$ (deg) & 66.5 & 2.5 & 69.0 & 64.0\\
    &$E_{\rm d}$ (keV)     &26.6   & 0.8 &26.7  & 26.6 \\
    & $\gamma~^{\rm b}$ & (0.25) &  --- & (0.25)  & (0.25) \\
    & $\Delta \psi$ (deg) $^{\rm b}$ & (180) &   ---  & (180)   & (180)\\\hline  
  \multicolumn{2}{l}{$S(E)$}\\
    &$R$ (deg/keV)          &2.2 & 0.5 & 3.2 & 2.6\\
    &$E_{\rm piv}$ (keV)    &62 & 15 & 72  & 76\\
\hline  
  \multicolumn{2}{l}{$T$ (ks)}    &  38.0    & 2.0   &  38.0  & 37.0\\
  \multicolumn{2}{l}{$A_0$ (sec)}   &  0.59   & 0.03 &  0.47  & 0.46  \\
  \multicolumn{2}{l}{$\psi_0$ (deg)} & 150   &  20   &   150   & 150\\
\hline  
 \multicolumn{2}{l}{$Z_4^2$} & 59.70  & -- &  59.57 & 62.47\\
 \hline \hline 
 \end{tabular}
 \label{tbl:EDPV_bestpara}
 \smallskip
 \begin{itemize} 
 \setlength{\baselineskip}{4mm}
  \item[$^{\rm a}$]: These errors apply to equations (\ref{eq:A(E)}), (\ref{eq:A'(E)}), and (\ref{eq:EDPV_square}).
 \item[$^{\rm b}$]:  Fixed base on some pilot studies.
 \item[$^{\rm c}$]:  With $\sigma=30$ fixed.
  \end{itemize}
 \end{center}
  \end{table}  

\begin{figure*}
\centerline{\includegraphics[width=120mm]{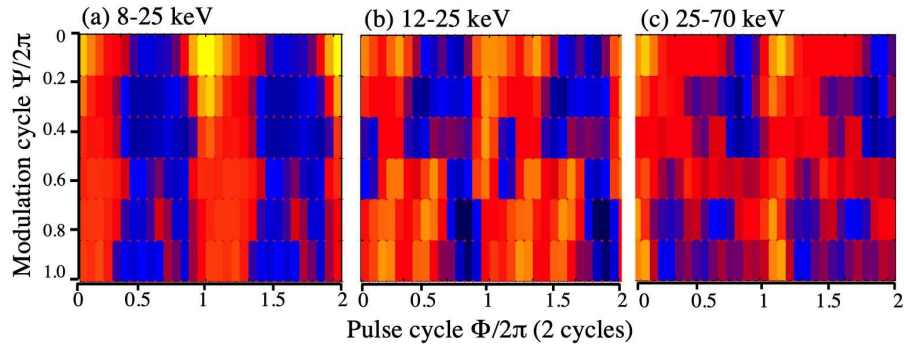}}
\caption{The  double-folded maps 
obtained after correcting all the photon arrival times 
using the EDPV parameters in Table~\ref{tbl:EDPV_bestpara},
together with $A_0=0.52$ s, $ \psi_0=190^\circ$, 
and $T=36$ ks.
Panels (a), (b), and (c) should be compared with
panels (b), (e), and (f) in Fig.~\ref{fig8:dblfld_6bands}, respectively.}
\label{fig13:dblfld_Lo8}
 \end{figure*}

Figure~\ref{fig13:dblfld_Lo8} shows double-folded maps produced
after correcting the arrival times of all photons for the EDPV effects,
using the final solution  in Table~\ref{tbl:EDPV_bestpara}.
Compared  with  the 0th-stage results in Fig.~\ref{fig8:dblfld_6bands},
the 8--25 keV map shows a more  straight ridge.
Although this is simply the 1st-stage effect,
the benefit of the 2nd-stage corrections emerges
in the two higher-energy maps,
where the main pulse peak appears at a constant 
pulse phase of $\Phi/2\pi \sim 1.0$, like in 8--25 keV.
In addition, the 12--25 keV map reveals two sub-pulse loci
which are also rather straight.
Thus,  Fig.~\ref{fig13:dblfld_Lo8}  illustrates what can, 
and what cannot,  be achieved in the 2nd stage.
Namely, our EDPV corrections apply
an energy dependent horizontal displacement 
to each row of the  map,
to make the pulse ridge(s)  vertically as straight as possible.
In contrast, details of the $\Psi$-dependent variations in the pulse profiles
cannot be fixed by this method.

\begin{figure}
\centerline{\includegraphics[width=85mm]{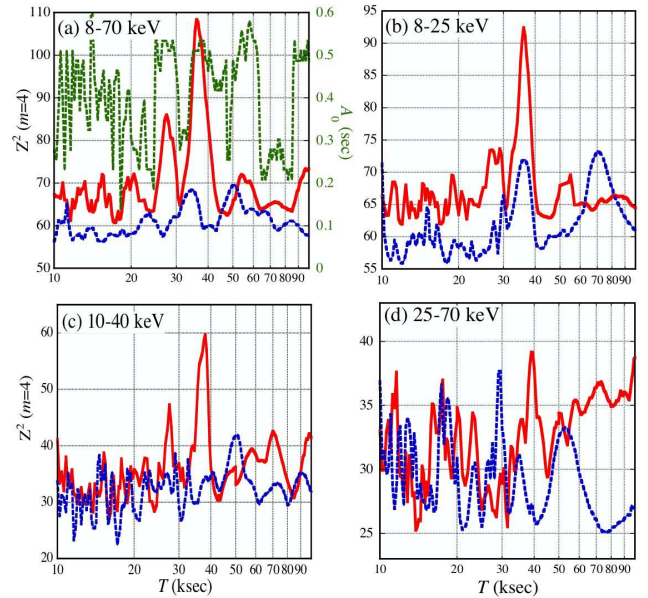}}
\caption{The $m=4$ DeMDs in the four energy intervals.
Solid red lines indicate the results obtained in the 2nd stage ({\it i.e.}, the EDPV corrections),
using the parameters  in Table~\ref{tbl:EDPV_bestpara}.
Dashed blue curves show the 1st-stage results, {\it i.e.}, 
the simple demodulation using equation~(\ref{eq:modulation});
the blue one in (b) is identical to the $m=4$ result
in Fig.~\ref{fig7:Tscan_8-25keV}.
In (a), the behaviour of $A_0$ in the 2nd stage is shown in green.
See text for further details.
 }
\label{fig14:EDPV_Tscan}
 \end{figure}

\begin{figure}
\centerline{
\includegraphics[width=83mm]{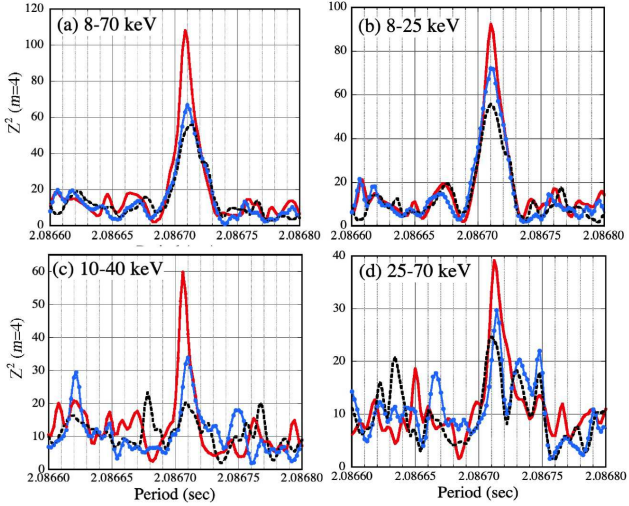}
}
\caption{The pulse periodograms in the four energy ranges.
The meanings of the  red  and  blue curves 
are the same as in Fig.~\ref{fig14:EDPV_Tscan},
wheres the dashed black lines represent the results
without any demodulation correction (0th stage).
 }
\label{fig15:EDPV_PG_3cases}
\end{figure}

\begin{figure*}
\centerline{
\includegraphics[width=15cm]{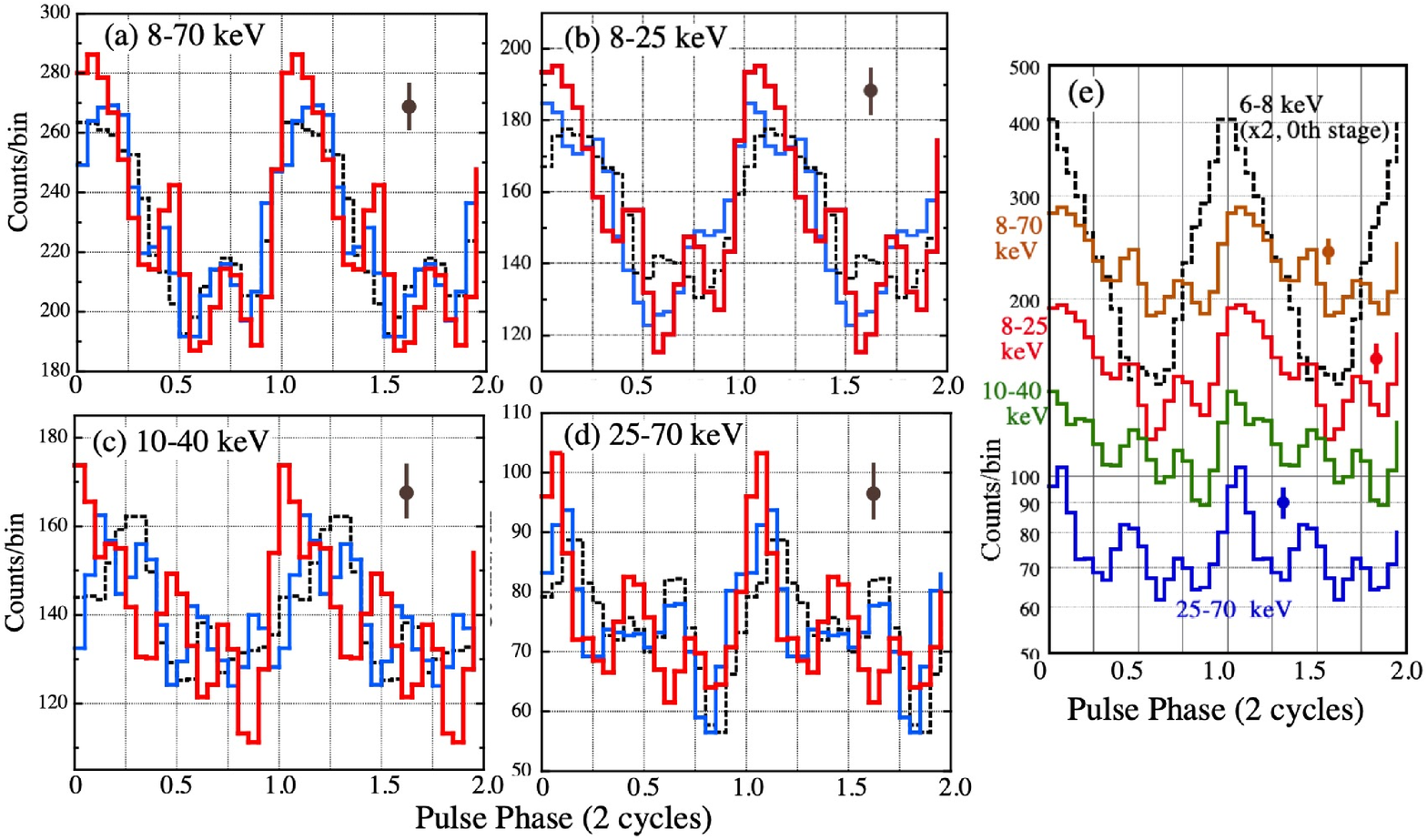}
}
\caption{
(a) to (d): Pulse profiles in the four energy ranges,
folded using the parameters at the peak of the corresponding
periodogram in Fig.~\ref{fig15:EDPV_PG_3cases}.
The meanings of the solid red, solid blue, and dashed black
lines are the same as in Fig.~\ref{fig15:EDPV_PG_3cases}.
(e) A compilation of the red profiles in panels (a) through (d),
shown together using a logarithmic ordinate.
The 10-40 keV profile was scaled by 0.8 just to avoid overlap.
The dotted black curve shows the  6-8 keV profile 
(doubled for presentation),
obtained with no demodulation.
 }
\label{fig16:EDPV_Pr_3cases}
\end{figure*}

For more quantitative evaluations of the EDPV corrections, 
we conducted a series of  studies 
using the following four energy intervals;
(a) the overall 8--70 keV band,
(b) the 8--25 keV range  where the analysis started,
(c) the 10--40 keV band where he EDPV parameters are optimized,
and (d) 25--70 keV to test  the recipe against the hardest photons.
These energy ranges mutually overlap,
and none of (a), (b), or (d) is disjoint from (c)
which was used for the parameter optimisation.
The derived results are presented in 
Fig.~\ref{fig14:EDPV_Tscan}, 
Fig.~\ref{fig15:EDPV_PG_3cases}, 
and Fig.~\ref{fig16:EDPV_Pr_3cases},
where  panels (a) through (d) represent
the four energy bands,  respectively.
The EDPV parameters were all fixed throughout
to those  in Table~\ref{tbl:EDPV_bestpara},
and $m=4$ was retained as before.

Figure~\ref{fig14:EDPV_Tscan} presents  the DeMDs 
derived in the  2nd stage (solid red),
in comparison with those from the 1st-stage analysis (dashed blue).
As $T$ is varied, both $A_0$ and $\psi_0$
(or $A$ and $\psi$) were allowed to vary,
and $P$ to change over the error range of equation~(\ref{eq:P0}).
Figure~\ref{fig15:EDPV_PG_3cases} compares the pulse periodograms
in each energy range, derived under three conditions;
without demodulation (0th stage; dashed black),
with the energy-independent demodulation (1st stage; solid blue),
and incorporating the EDPV corrections (2nd stage; solid red).
The red and blue curves were computed using the parameters 
at the $T\sim  36$ ks peak in Fig.~\ref{fig14:EDPV_Tscan}, 
and the associated values of $A_0$ and $\psi_0$ 
(or $A$ and $\psi$).
The parameters of these periodogram peaks 
are summarised in Table~\ref{tbl:Z2_summary},
separately for the 3 conditions.
Finally, Fig.~\ref{fig16:EDPV_Pr_3cases} presents the folded pulse profiles,
derived under the same  3  conditions.
Figure~\ref{fig16:EDPV_Pr_3cases}e compiles the finally derived 4 pulse profiles, 
and the 6-8 keV one from the simple folding.

The three figures reveal impacts of the EDPV corrections,
which can be summarised in the following  points.
\begin{enumerate}
\vspace*{-2mm}
\item 
In Fig.~\ref{fig14:EDPV_Tscan},
the 36 ks peak appeared strongly in all  2nd-stage DeMDs,
particularly  in (a) through (c),
and the $T\sim 72$ ks peak disappeared in (b).
Even in (d), the peak appears at $T=39.0$ ks,
which is still within 1.3  sigmas of equation~(\ref{eq:demodpara_8-25keV}).
\item
The values of $T$ via the EDMP corrections (Table~1)
are generally consistent with  equation~(\ref{eq:demodpara_8-25keV}),
and the \Su\ measurement in equation~(\ref{eq:36ks_Suzaku}).
The periodogram peak remains at $P_0$
of equation~(\ref{eq:P0}) within the error.
Furthermore, $A_0 \sim 0.52$ sec and $\psi_0 = 180^\circ \pm 30^\circ$
obtained in the 2nd stage (Table~\ref{tbl:Z2_summary})
are both relatively energy independent.
\item
In all the four energy bands,
the EDPV corrections  increased the pulse significance 
by $\delta Z_4^2>8$, compared to those from the 1st stages
(Table~\ref{tbl:Z2_summary}). In 8--70 keV, {\it e.g.},
the increase from $Z_4^2=68.43$ (1st stage) 
to $108.38$ (2nd) by $\delta Z_4^2=39.95$
means  an 8-orders-of-magnitude decrease 
in the pre-trial probability (Appendix A),
even though that in the post-trial probability is not obvious
because we must consider the number of independent  trials 
in optimizing the 7+2 EDPV parameters.
(This cannot be easily Monte-Carlo evaluated,
as the  parameter scanning demands
huge computational times.)
\item
In response to the increase in $Z_4^2$ 
(equation \ref{eq:PF_vs_Z2}),
the PF increased 
to $\gtrsim 18\%$  in all energies (Table~\ref{tbl:Z2_summary}).
Even though these are still lower than measured in $<8$ keV,
the puzzle with Fig.~\ref{fig5:LDUD} has  been solved
at least partially.
\item
Figure~\ref{fig16:EDPV_Pr_3cases}e provides the most
impressive result of the present study,
where the pulse profiles have become very similar and 
in-phase to one another, with a sharp rise and a slow decline.
The main peak is at $\Phi/2\pi \sim 0.1$, 
and two sub-peaks 
are seen at $\Phi/2\pi  \sim 0.4$ and $\sim0.7$.
The  separations among these peaks, 1/4--1/3 pulse cycles,
justify our use of $m=4$.
These features sharpen towards higher energies;
so does the periodogram peak.
In contrast to the  irregular energy dependence  in Fig.~\ref{fig4:Pr},
the regular pulse profiles revealed here 
highlight the achievements in the 2nd stage.
\end{enumerate}

Of course, these results might be trivial for the 10--40 keV interval (panels c),
because the EDPV parameters have been optimized
to enhance the pulse visibility there.
To remove  this concern,
we conducted the same parameter optimisation 
(except $\Delta \psi$ and $\gamma$ which are fixed)
using the other three energy ranges. 
Then, in any of them, 
the optimum values of the 7 parameters became consistent, 
within errors, with those determined in 10--40 keV 
(Table~\ref{tbl:EDPV_bestpara}).
Equivalently,  in none of  (a), (b), or (d),
the parameter re-adjustment  increased $Z_4^2$  by more than $\sim 2$, 
above that  (``EDPV'' row in Table~\ref{tbl:Z2_summary})
specified by the parameters determined in 10--40 keV.
In particular, the agreement between the  two disjoint intervals, 
8--25 keV and 25--70 keV, is assuring.

In closing this section,
we may  examine how  the three energy-dependent functions contribute 
individually to the $Z_4^2$ increment.
Although the answer depends on the energy,
generally  replacing $A$ to $\tilde{A}(E)$ is most effective,
contributing about half of $\delta Z_4^2$ 
above the case of simple demodulation.
Then, the remaining half is contributed  about equally 
by $\tilde{\psi}(E)$ and $S(E)$.
The three corrections are all mandatory.

\section{Discussion}
\label{sec:discussion}

\subsection{Summary of the obtained results}
\label{subsec:summary_of_results}
We analysed the \NuS\  data of  \oneE\ 
acquired on 2016 August 23 to 24,
mainly focusing on its pulsation
detected at the period of equation (\ref{eq:P0}).
As reported by CZ20 who already analysed the same data,
the PF decreased markedly in the 
intermediate energy range  (Fig.~\ref{fig5:LDUD}),
and the folded pulse profiles exhibited
complicated energy dependences (Fig.~\ref{fig4:Pr}).
Using these puzzles as a springboard,
we carried out further timing analysis in two stages.

In the 1st stage,
we conducted the demodulation analysis,
employing  equation~(\ref{eq:modulation}) and assuming 
that $A$ and $\psi$ are both constant 
over the data-accumulation energy range.
In $< 8$ keV where the SXC dominates, 
the pulse was free from phase modulation, like in the \Su\  data.
In the 8--25 keV band,  we  reconfirmed 
the 36 ks pulse-phase modulation (Fig.~\ref{fig7:Tscan_8-25keV})
discovered with \Su.
The detection of this effect  both in an outburst and  quiescence
suggests that it is a persistent phenomenon,
presumably due to celestial mechanics,
rather than a transient  episode in a high activity.
This phenomenon is also regarded as rather common to magnetars,
because it has been confirmed in two contrasting sources,
\oneE\ with the fastest rotation and high variability,
and 4U~0142+61 which is relatively persistent
and has two orders of magnitude larger characteristic age.
We hence follow Paper I, 
and  interpret the 36 ks periodicity as the slip period 
associated with free precession of the NS,
which is axially deformed to $\epsilon \sim 0.6 \times 10^{-4}$.

In $> 10$ keV, in contrast,
the demodulation was effective only to a limited degree;
the DeMDs did not show noticeable peaks at $T \sim 36$ ks
(Fig.~\ref{fig10:Tscan_above_10keV}a),
and the  puzzles have remained unsolved.
Through detailed inspections of the data 
(Fig.~\ref{fig10:Tscan_above_10keV}b, c),
we recognized that the problem stems from the strong energy dependence (EDPV) 
of the HXC pulse properties,
and carried out the 2nd-stage analysis,
for the first time in our studies of this subject.
That is, guided by the inspections and further data analysis
 (Fig.~\ref{fig11:EDPV_guess}),
we modeled the EDPV effects 
by the three empirical functions, 
$\tilde{A}(E)$, $\tilde{\psi}(E)$, and $S(E)$.
By optimizing their parameters  in the 10--40 keV range,
and using the results to correct the photon arrival times
in energy-dependent ways (Fig.~\ref{fig12:EDPV_bestfit_curves}),
the 36 ks peak in the DeMDs have been restored,
from 10 keV to 70 keV (Fig.~\ref{fig14:EDPV_Tscan}).
This establishes that the 36 ks pulse-phase modulation 
is an intrinsic  property of the HXC of this object.
The corrections have also solved the puzzling behaviour of the HXC pulses;
the PF  has increased to $\gtrsim 18\%$,
and the irregular energy dependence of the pulse profiles 
has been rectified  (Fig.~\ref{fig16:EDPV_Pr_3cases}).

In several places,
we used the running average.
As describe in Appendix B, this process suppresses 
noise with high spatial frequencies,
to  make the pulse profiles and double-folded maps easier to grasp.
Although it also affects the data statistics,
the major quantitative evaluations of the present work remain intact,
because they are all based on the $Z_n^2$ method
that incorporates no running average.

Below, we examine these 2nd-stage results from three aspects;
appropriateness of this EDPV modeling,
geometrical scenarios in terms of the  free precession,
and  possible astrophysical interpretations.

\subsection{Reality of the EDPV modeling}
\label{subsec:exam_EDPV_modeling}

Our EDPV modeling
is apparently supported by the two results
described in \S~\ref{subsubsec:EDPV_results};
the different portions of the 8--70 keV energy band
yielded  consistent  EDPV parameters,
and the  pulse phase has become highly coherent
as a function of energy (Fig.~\ref{fig16:EDPV_Pr_3cases}e).
Nevertheless,  the modeling is admittedly very extraordinary;
equation (\ref{eq:A(E)}) claims
that $A$ is enhanced at $\sim 22$ keV,
almost like a resonance, to reach  $\sim \pm 1/4$ of a pulse cycle.
The phase reversal in $\Psi$ was also rather unexpected.
In addition, the data in  Fig.~\ref{fig11:EDPV_guess}a
are subject to large errors.
Yet another concern is
that  the several characteristic energies ({\it e.g.}, 8, 10, 21.6, and 26.6 keV)
involved in our modeling are apparently not
accompanied by any noticeable spectral features
(Fig.~\ref{fig1:spec};  Fig.~2 of CZ20).
Therefore, we should carefully evaluate the reality of this picture,
and examine other possible modelings.

Although the data points in Fig.~\ref{fig11:EDPV_guess}a
suggest  smooth variations of $A$, 
this can be an artifact due to the use of  overlapping data points.
The truth might be that  $A$  jumps abruptly,
at some energies, between $A \sim 0.1$ and $\sim 0.5$.
We can emulate such a case by modifying equation (\ref{eq:A(E)}) as
\begin{equation}
\tilde A'(E) = A_0 \left[  a_{\rm f} + 
\frac{1- a_{\rm f}}{   1.0+\left\{ (E-E_{\rm c})/E_{\rm w} \right\}^4  } \right]~.
\label{eq:A'(E)}
\end{equation}
Here, the 4th power expresses more rapid changes
than the original Lorentzian, 
using the same three parameters as before.
This $\tilde{A}'(E)$, together with the same $\tilde{\psi}(E)$ and $S(E)$ as before,
were substituted into equation (\ref{eq:EDPV}), 
and again all parameters of $\tilde{A}'(E)$, $\tilde{\psi}(E)$, 
and $S(E)$ were optimized.
The  results are summarised in Table~\ref{tbl:EDPV_bestpara},
and the profile of the optimum
$\tilde{A}'(E)$ is drawn in Fig.~\ref{fig17:A_and_A'}
in comparison with the original $\tilde{A}(E)$.
Indeed, $\tilde{A}'(E)$ emulates a transition-like behaviour
between $A \sim 0.1$ and $A \sim 0.45$,
and reconfirms the marked increase of $A$ at $\sim 20$ keV.
However, the maximum  $Z_4^2$  thus achieved  was lower by 0.13, 
than that obtained with  equation~(\ref{eq:A(E)}).
Furthermore,  when the parameters are optimized  in the 8--70 keV range,
$\tilde{A}'(E)$ again gave a value of $Z_4^2$ 
which is  lower by 3.75 than $\tilde{A}(E)$.
Therefore, we do not find a good reason 
to replace $\tilde{A}(E)$  with $\tilde{A}'(E)$.

\begin{figure}
\begin{center}
\includegraphics[width=4.3cm]{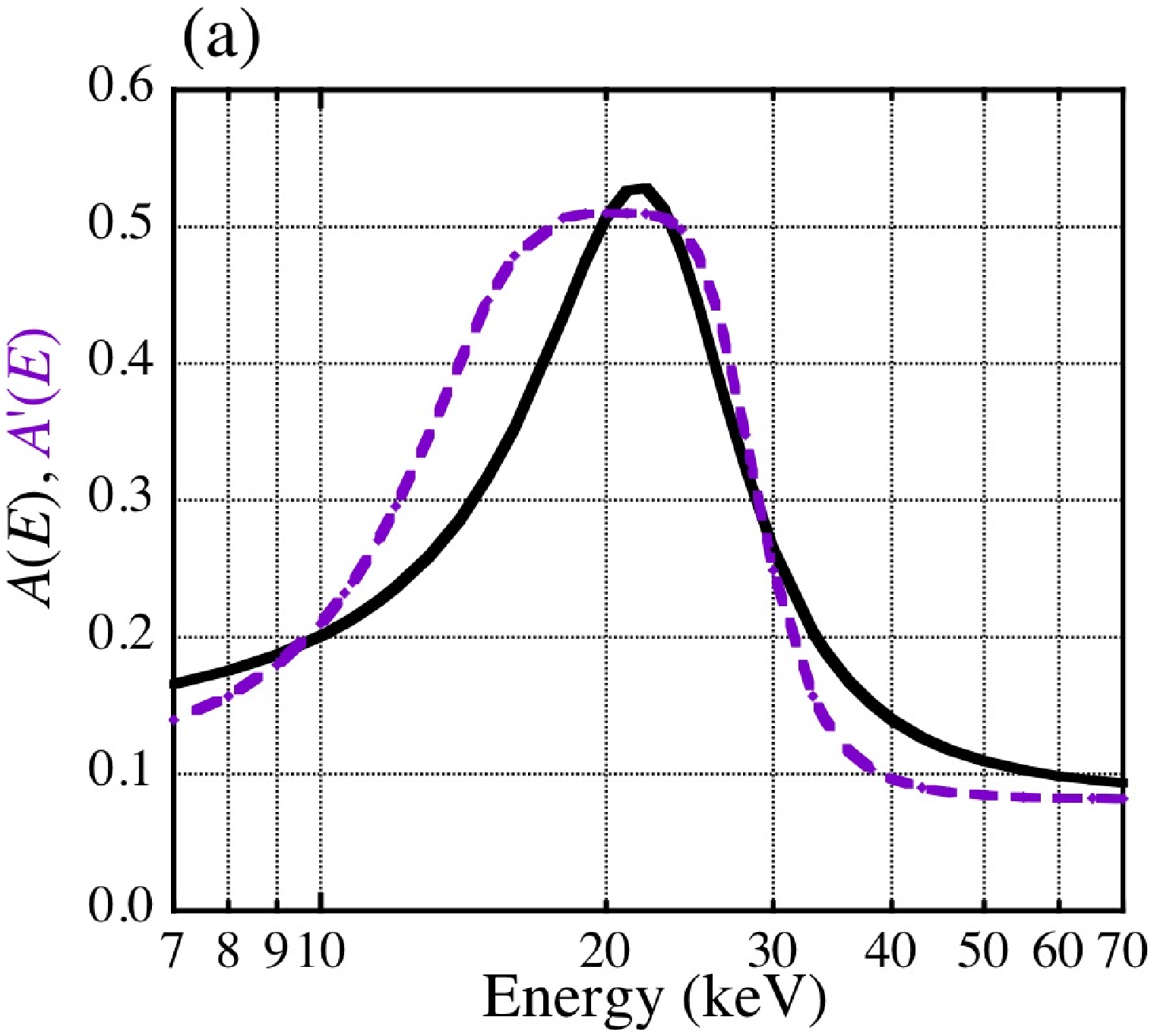}
\includegraphics[width=4.0cm]{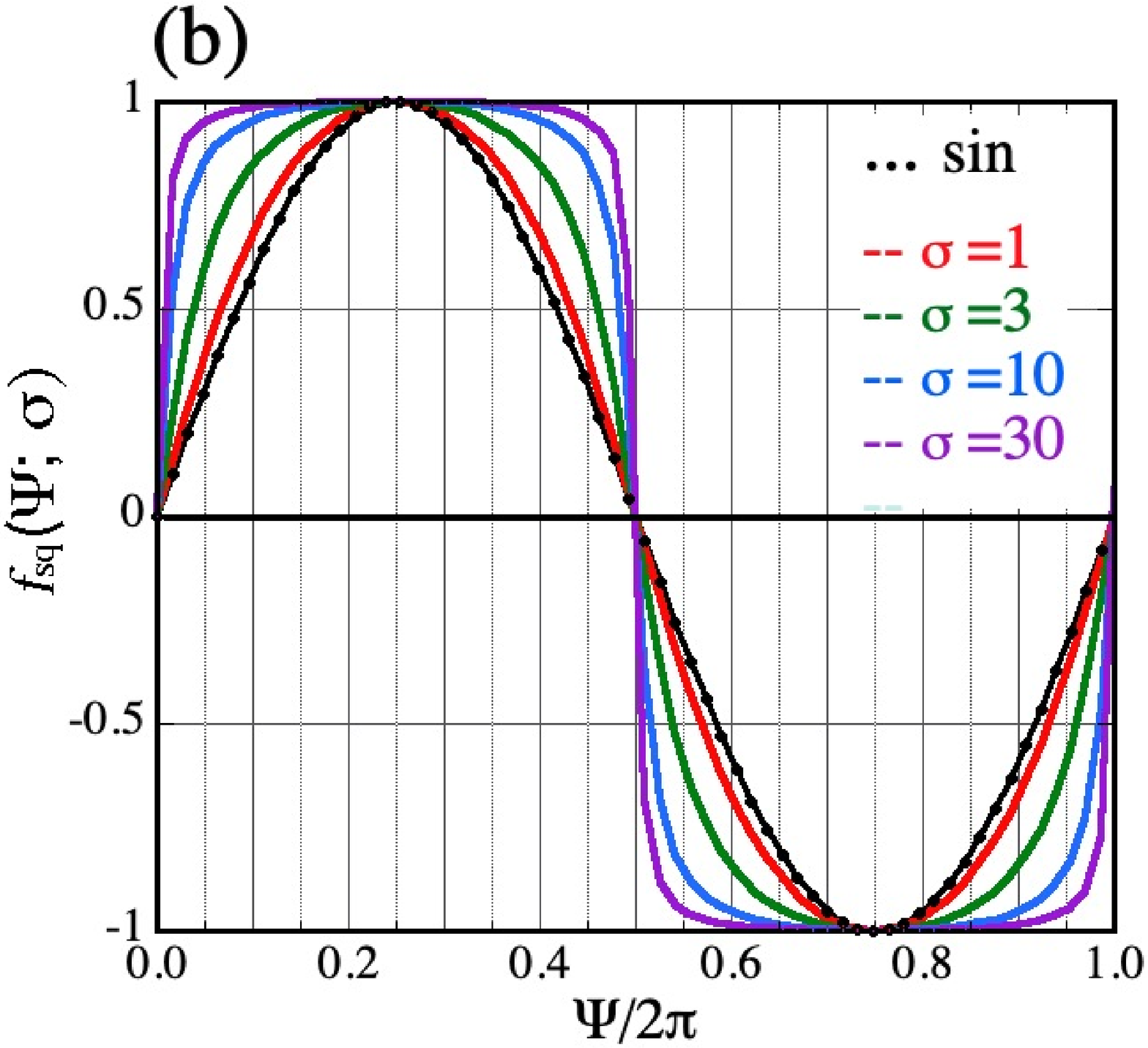}
\end{center}
\caption{
(a) The behaviour of $\tilde{A}'(E)$ (equation~\ref{eq:A'(E)}) in dashed purple,
compared with that of $\tilde{A}(E)$ (equation~{\ref{eq:A(E)}) in solid black.
Both have been optimized in the 10--40 keV range.
The corresponding parameters are given in Table \ref{tbl:EDPV_bestpara}.
(b) The square-wave-like modulation profiles $f_{\rm sq}(\Psi; \sigma)$
specified by equation~(\ref{eq:EDPV_square}),
shown against $\Psi/2\pi$ for different values of $\sigma$.
}
 }
\label{fig17:A_and_A'}
\end{figure}

Returning to the original Lorentzian form, 
the model-implied value of $A$ at $E\sim E_{\rm c}$ is similar 
to that found with \Su\  (Paper I), $A=0.52 \pm 0.14$ sec
(although in that case $A$  was not strongly energy dependent at $E \gtrsim 10$ keV).
Such a large $A$ could arise (Paper I)
if the pulse profile comprises two peaks separated by about half a pulse cycle,
and their relative intensities  interchange through the 36 ks phase.
If so, we expect the pulse peak to draw
 a square-wave like locus on the double-folded map,
rather than the sinusoidal variation assumed  so far.
Actually, in  Fig.~\ref{fig8:dblfld_6bands}e,
the 12--25 keV pulse peak appears to keep
a rather constant phase over $\Psi/2\pi=0-0.5$,
and then jumps to a different phase} over $\Psi/2\pi=0.5-1.0$.
Introducing another parameter $\sigma \geq 0$,
we hence modified the $\sin (\Psi)$ factor
(for simplicity $\tilde{\psi}$ was set 0)
 in equation (\ref{eq:EDPV}) to \citep{Makishima16}
\begin{equation}
f_{\rm sq} (\Psi; \sigma) \equiv \frac{{\arctan} \left\{\sigma \sin (\Psi) \right\} }
{{\arctan}(\sigma)}~,
\label{eq:EDPV_square}
\end{equation}
which reduces to $\sin (\Psi)$ if $\sigma \ll 1$.
As shown in Fig.~\ref{fig17:A_and_A'}b,
$f_{\rm sq} (\Psi; \sigma)$ becomes saturated as $\sigma$ increases,
and approaches a square wave 
that oscillates between $ \pm 1$  with 50\% duty ratio.
Replacing $\sin (\Psi)$ in  equation~(\ref{eq:EDPV}) with $f_{\rm sq} (\Psi; \sigma)$,
we repeated the 2nd-stage analysis of the 10--40 keV data.
As $\sigma$ is changed,
all the other model parameters were re-optimized.
Then, as in the last column of Table~\ref{tbl:EDPV_bestpara},
$Z_4^2$ increased by 2.77 for $\sigma \sim 30$ 
(with a typical uncertainty of $\pm 20$)
over  the sinusoidal case ($\sigma \rightarrow 0$).
When selecting the 8--70 keV range instead,
the case with $\sigma \sim 30$ was again favored,
as  $Z_4^2$ was larger by $\sim 2.0$.
Therefore, as suggested by Fig.~\ref{fig8:dblfld_6bands}e,
the actual phase-modulation waveform could be fairly square-wave like,
rather than sinusoidal.
However, the data preference for $f_{\rm sq} (\Psi; \sigma)$
than $\sin (\Psi)$ may not be obvious statistically,
because the introduction of $\sigma$ means
an additional increase in the number of trials.

From these evaluations, we conclude that 
the functional forms of $\tilde{A}(E)$, $\tilde{\psi}(E)$,  and $S(E)$, 
expressed by equations (\ref{eq:A(E)}),  (\ref{eq:psi(E)}), and  (\ref{eq:S(E)}) respectively,
provide a reasonable account of the pulse-phase behaviour above 8 keV,
although the modulation waveform as a function of $\Psi$ 
may deviate  somewhat from sinusoidal to become square-wave like.
In any case, these results must be regarded as tentative,
possibly with considerable room for future improvements.

\subsection{Possible emission geometry}
\label{subsec:emission_geometry}

\begin{figure}
\begin{center}
\includegraphics[width=4.2cm]{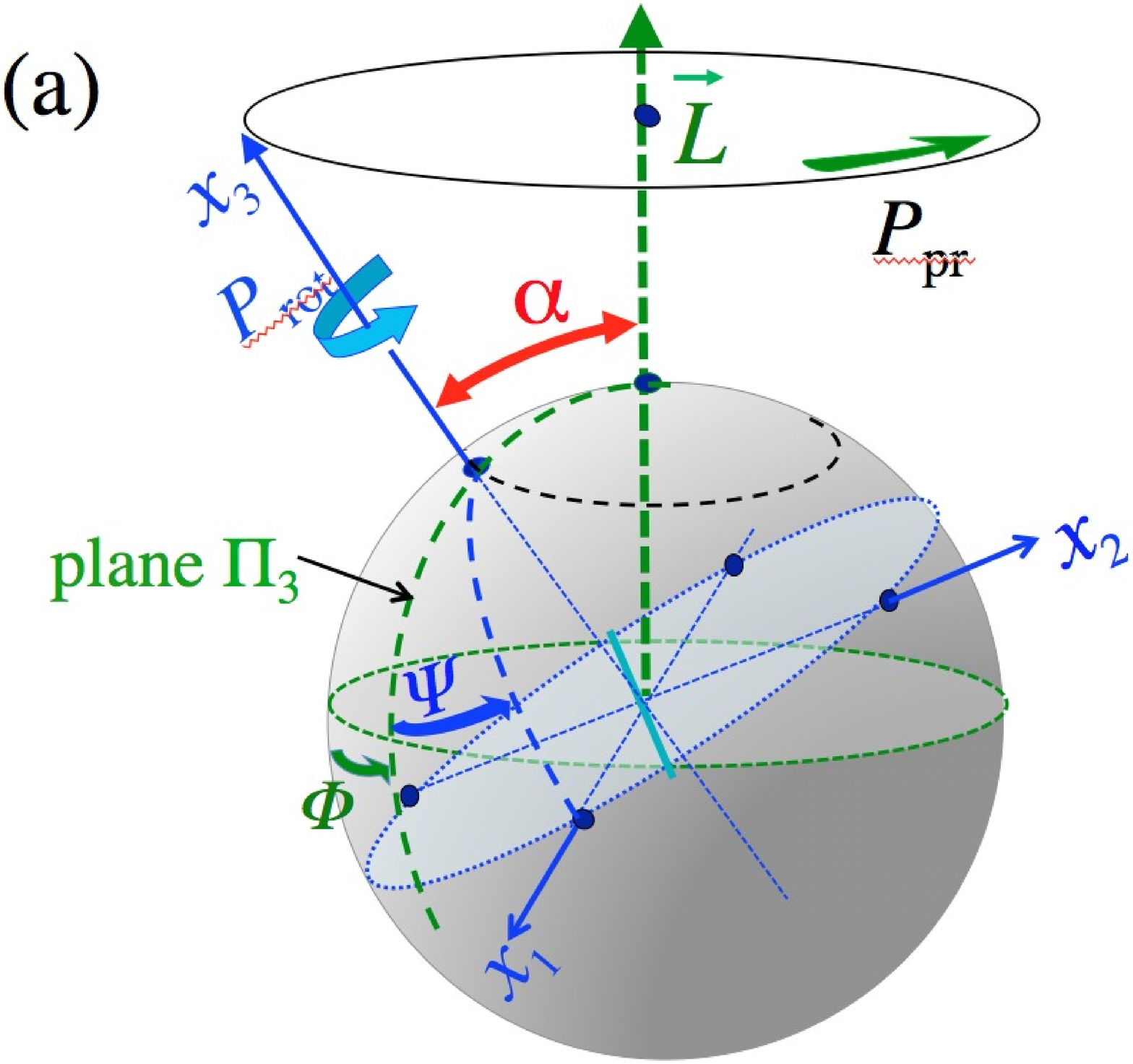}
\includegraphics[width=4.1cm]{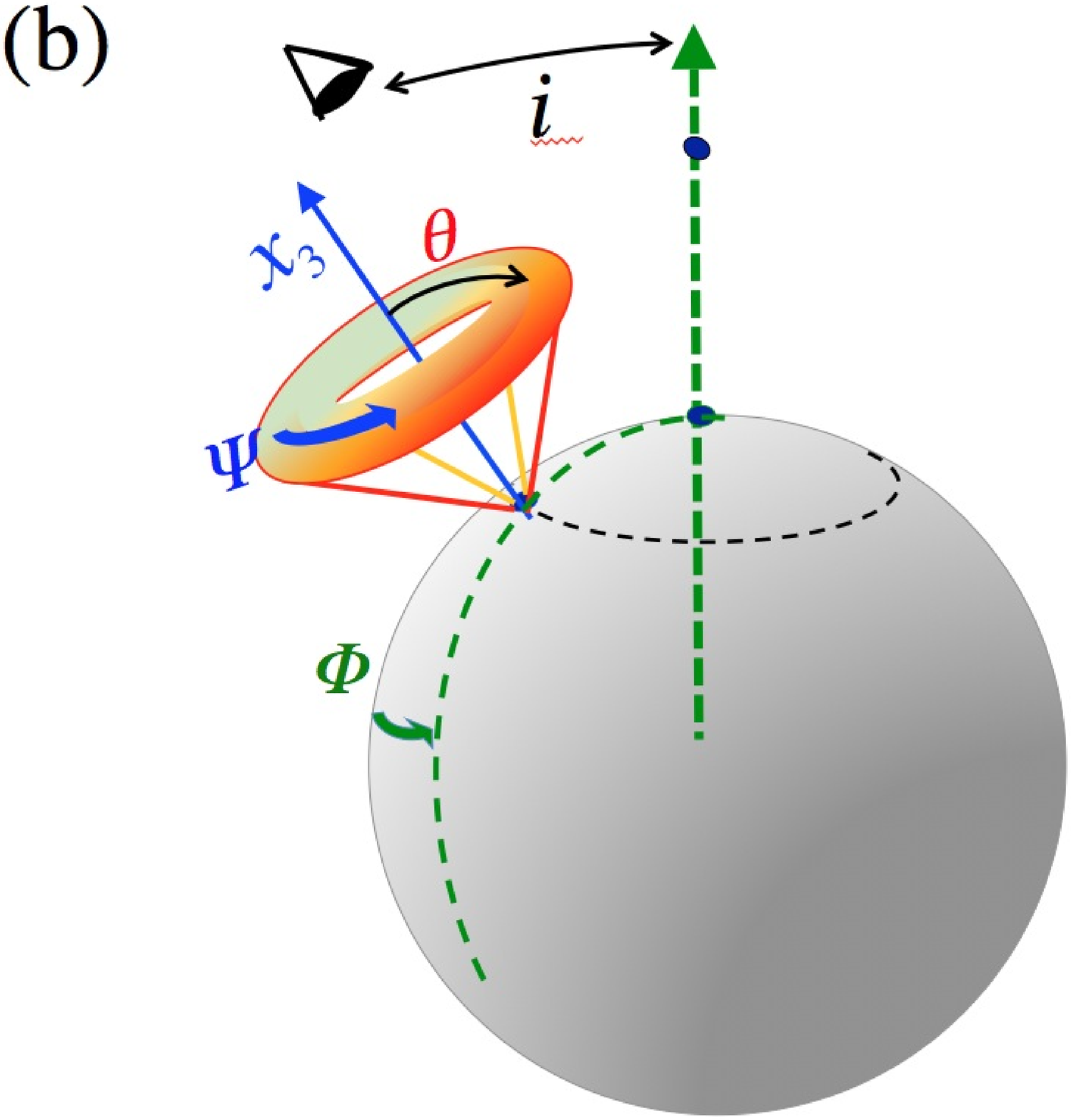}
\end{center}
\caption{(a) The assumed  geometry of the object, 
where the observer and the star's centre are on the plane of the sheet.
(b) The assumed conical beam pattern of the HXC,
where the azimuthal variation of the emissivity is indicated by colors.
}
\label{fig18:geometry}
\end{figure}

We next  consider geometrical models 
that can explain  the observed  pulse-phase behaviour,
in  the framework
that  the 36 ks period  represents the slip period $T$
of equation~(\ref{eq:slip}),
associated with the free precession of the NS.
Below, we refer to Fig.~\ref{fig18:geometry}a,
which utilises the nomenclature 
introduced in \S~\ref{sec:intro}.
Here,  $\hat{x}_3$  is identified with the stellar dipole-field axis,
$\Pi_3$ denotes the plane defined by $\hat{x}_3$  and $\vec{L}$,
and the observer's line of sight  lies on the plane of the sheet.
Our basic assumption is 
that the HXC emissivity pattern is energy dependent,
but not time dependent if expressed
in the $(\hat{x}_1, \hat{x}_2, \hat{x}_3)$ frame 
which is fixed to the NS:
the observed time variability is solely due to the motion 
of the stellar frame,
relative to the inertial frame
where the observer is located.

With this setup, 
the basic idea to explain the periodic pulse-phase modulation
in a magnetar is to invoke three levels of symmetry breaking
(\S~1; Paper I; \citet{Makishima14, Makishima19});
(i) $\alpha \ne 0$, as evidenced by the clear pulsation;
(ii) $\epsilon \ne 0$, {\it i.e.}, the NS is aspherical (but axially symmetric);
and (iii) the HXC emission pattern is asymmetric around $\hat{x}_3$.
According to  (i), $\Pi_3$ rotates around $\vec{L}$ with the period $P_{\rm pr}$,
and its rotation phase relative to the observer's line of sight
is identified with the pulse phase  $\Phi$ of equation~(\ref{eq:pulse_phase}).
When  (ii) holds as well,
the NS also rotates around $\hat{x}_3$ {\it relative to} $\Pi_3$,
with the slip period $T$.
The angle  of this  rotation seen from $\Pi_3$ coincides with 
the modulation phase $\Psi$ defined by equation~(\ref{eq:modulation_phase}).
The  triplet $(\Phi, \alpha, \Psi)$  serves as the three Euler angles
that transform  the observer's frame
to $(\hat{x}_1, \hat{x}_2, \hat{x}_3)$.
On a double-folded map drawn on the $(\Phi/2\pi, \Psi/2\pi)$ plane,
the X-ray intensity depends on $\Phi$, and also becomes dependent on  $\Psi$
when the condition (iii) sets in ({\it e.g.}, Fig.~\ref{fig8:dblfld_6bands}b).

The condition (iii) is fulfilled \citep{Makishima14}
if the emission region is
displaced from $\hat{x}_3$ on the NS surface,
or the  beaming direction is tilted from $\hat{x}_3$.
Here, we employ the latter case, and assume, 
with some modifications over the toy model of  \citet{Makishima19},
that the emission has a conical beam pattern around $\hat{x}_3$
as illustrated in Fig.~\ref{fig18:geometry}b.
The cone is assumed to be geometrically symmetric around $\hat{x}_3$, 
with a half opening angle $\theta$,
and emit hard X-rays along its generatrices,
with a full-width-at-half-maximum divergence angle of 
$\pm 45^\circ$ as measured from the cone surface.
The case with $\theta \rightarrow 0$ reduces to 
the pencil-beam configuration aligned with $\hat{x}_3$,
whereas $\theta \rightarrow 90^\circ$
represents the fan-beam configuration
as assumed in Fig.12 of Paper I to explain the \Su\ result.
Moreover, the cone is assumed to be physically asymmetric, 
so that  its directional emissivity depends on the cone's azimuth  
(the color gradient in Fig.~\ref{fig18:geometry}b) 
as $\propto \left[1 + a \cos \left\{\Psi-\tilde{\psi}(E) \right\} \right]$,
where $a~(0\le 1 \le 1)$ specifies the degree of asymmetry around $\hat{x}_3$.
The condition (iii) requires both $\theta >0$ and $a>0$.

\begin{figure*}
\begin{center}
\includegraphics[width=16cm]{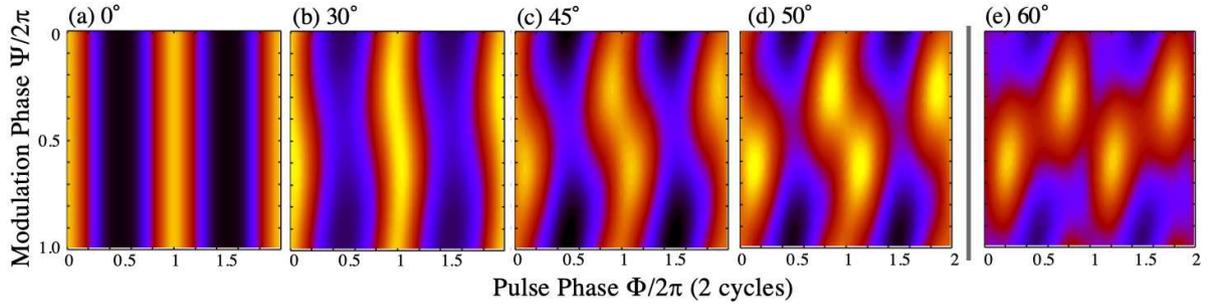}
\end{center}
\caption{The expected X-ray intensity (corlor),
calculated assuming a toy model (see text for details)
for several values of $\theta$.
The coordinates, $(\Phi/2\pi,\Psi/2\pi)$,
are the same as in Fig.~\ref{fig8:dblfld_6bands}
and Fig.~\ref{fig12:EDPV_bestfit_curves}.
Panel (e) may be appropriate for the \Su\ result in Paper I.
}
\label{fig19:fanbeam_simulations}
\end{figure*}

Figure~\ref{fig19:fanbeam_simulations} shows 
the pulse-phase behaviour,
predicted by the above toy model
for several representative values of $\theta$.
We assumed a viewing inclination angle $i=40^\circ$ to $\vec{L}$,
and $\alpha = 45^\circ$
to avoid  self-occultation effects.
In addition, we assumed $a=0.5$ (a mild asymmetry), 
and $\tilde{\psi}=170^\circ$ ignoring 
its complex energy dependence. 
The emission from the other pole was neglected,
as well as  general relativistic light bending effects.
The calculated modulation amplitude thus increases as $\theta$ gets larger,
up to $A \sim \pm P_0/4$ at $\theta \gtrsim 50^\circ$.
Therefore, the behaviour of $\tilde A(E)$ can  be explained
if $\theta$ somehow increases
 to $\theta \sim 50^\circ$ at $E\sim 22$ keV.
In particular, Fig.~\ref{fig19:fanbeam_simulations}d looks
rather similar to Fig.~\ref{fig8:dblfld_6bands}e.
Panel (e) with $\theta = 60^\circ$ may apply to the \Su\ result,
wherein two comparable pulse peaks (from the same magnetic pole),
half a cycle apart,
interchanged in intensity as a function of $\Psi$ (Fig. 2b of Paper I).
Another effect of increasing $\theta$ is 
that the modulation waveform gets gradually deviated 
from a sinusoidal shape, and becomes square-wave like;
this may agree with the data preference for equation~(\ref{eq:EDPV_square})
(Table~\ref{tbl:EDPV_bestpara} last column).

The above toy model  thus works at least to a crude approximation,
but the results should be taken with several reservations.
First, Fig.~\ref{fig19:fanbeam_simulations} is meant to show
that some emission geometry can roughly explain the present observation,
rather than to claim that the selected model 
or the geometrical parameters are correct.
Second, the behaviour of $\tilde \psi(E)$ would also be 
incorporated into Fig.~\ref{fig19:fanbeam_simulations},
although the present calculation
took into account neither this  adjustment,
nor the $\sim 27$ keV reversal in $\Psi$.
Third, a similar sequence as changing $\theta$ can be 
obtained by fixing $\theta \sim 50^\circ$, 
and instead increasing $a$ from 0 to 0.5.
Fourth,  the behaviour of $S(E)$ is more difficult,
than those of $\tilde{A}(E)$ and $\tilde{\psi}(E)$,
to reproduce by this model,
and probably we need to incorporate also energy-dependent 
displacements in the emission region.
Lastly,  Fig.~\ref{fig19:fanbeam_simulations} predicts
that the $\Phi$-averaged X-ray intensity depends on $\Psi$.
Such a behaviour is seen in Fig.~\ref{fig8:dblfld_6bands},
although it is not obvious whether it matches the calculation.

Evidently,  a more advanced data analysis would be to compare 
numerically the observational data as in Fig.~\ref{fig8:dblfld_6bands},
with  the geometrical predictions as in Fig.~\ref{fig19:fanbeam_simulations}.
This would enable us to see 
whether the model can quantitatively reproduce the observation,
and if so, to constrain the geometrical parameters,
$\alpha$, $i$, $\theta(E)$, and $a(E)$, 
as well as  $\tilde{\psi}(E)$.
This attempt will be our future study.

\subsection{Astrophysical interpretations}
\label{subsec:strophys_interpretation}
The final step of the present study is to  seek for possible 
astrophysical interpretations of the obtained results, 
including the geometrical model  constructed in \S~\ref{subsec:emission_geometry}.
The attempt will be very crude and speculative,
because the hard X-ray emission mechanism of magnetars itself is still unknown,
and the above geometrical model could be one of many possibilities.
Below,  we consider six specific queries;
[Q1] does the assumed conical emission pattern have any astrophysical basis;
[Q2] how the asymmetry in $\Psi$ can be produced;
[Q3] what caused  $A$ to be larger in the \Su\ observation on average than in that of \NuS;
[Q4] how we can explain the strong EDPV effects in the  \NuS\ data,
and their absence in the  \Su\ data;
[Q5] why the EDPV of the \NuS\ data started at $\gtrsim 8$ keV 
where the SXC still contributes significantly;
and [Q6] is the scenario capable of explaining 
the results from 4U~0142+61 as well.
Since the phenomena involved here are all specific to magnetars,
we  should answer these queries 
based on astrophysics of strong MFs,
with the least amount of  ad-hoc ideas
that are unrelated to this extreme environment.

\subsubsection{Photon splitting process}
One promising emission mechanism of the magnetar HXC assumes
that gamma-ray photons (including the 511 keV annihilation photons in particular),
somehow created near the magnetic poles  under the ultra-strong MFs,
propagate across the magnetosphere and repeatedly experience,
instead of the electron-positron pair creation,
a quantum-electro-dynamical process called {\it photon splitting};
a gamma-ray photon splits into two softer photons,
with the aid of the strong MF \citep[{\it e.g.},][]{PhotonSplitting, Enoto10b},
while conserving the photon energy.
The photons thus increase in number but degrade in energy,
to form the HXC with very hard spectral slopes.
If this process is operating, 
the emission region will naturally acquire a conical geometry
formed by the MFs,
although the local directional emissivity would be rather complex, 
depending on  the MF direction and the photon polarization.
If the cone is not hollow physically, 
we may superpose a series of cones with different $\theta$.
These affirmatively answer [Q1] at least qualitatively.

\subsubsection{Magnetic multipole contributions}
The cone would be symmetric around $\hat{x}_3$
if the MF has a pure dipole configuration.
However,  the  MF of a magnetar is usually considered 
to comprise tilted multipoles  \citep[{\it e.g.},][]{MultiPole02,Tiengo13, multipole_variation},
which are much stronger and no longer symmetric around $\hat{x}_3$.
In fact, the characteristic three-peak pulse profiles 
in Fig.~\ref{fig16:EDPV_Pr_3cases}e
are reminiscent of the 4-peak profile of SGR~1900+14 
during its Giant Flare in 1988,
which was taken as evidence for a multipolar geometry \citep{4peak_profile}.
Then, the local MF intensity $B$ can depend on $\Psi$.
This implies $a>0$,
because the photon-splitting cross section is thought 
to depend strongly on $B$ and the photon energy $E$ 
approximately as \citep{2photon_crossec}
\begin{equation}
\sigma_{\rm 2p} \propto E^5 B^6~.
\label{eq:splitting_probability}
\end{equation}
This will provide a possible answer to [Q2].
Incidentally, based on this dependence,
weaker $B$ would  lead to a harder HXC slope,
because  the gamma-ray source photons  would then
stop splitting at relatively high energies \citep{Enoto10b}.
This mechanism may  explain the observed scaling
that less active magnetars show harder HXC (\S~\ref{subsec:spectra}).

The intensity and configuration of a magnetar's MF  
are considered to vary considerably with time,
particularly when the activity changes \citep[{\it e.g.},][]{multipole_variation}.
Hence it would be natural to presume 
that the multipole component of \oneE\ was  
stronger during the outburst observation with \Su.
Then, equation (\ref{eq:splitting_probability}) predicts
that the photon splitting continued to larger distances from the magnetic poles,
where the MF opening angle ($\sim \theta$) is larger,
and that $a$ was also larger 
because of the enhanced multipole MFs.
This could explain [Q3], {\it i.e.}, why the modulation amplitude 
was  larger ($A \sim 0.5$ s) in the \Su\ data 
than in the \NuS\ data except at $E \sim E_{\rm c}$.

\subsubsection{Possible proton cyclotron resonance}
The most challenging issue is [Q4], namely,
how to interpret the EDPV effects which were seen only in the \NuS\ data.
As suggested by Fig.~\ref{fig5:LDUD} and Fig.~\ref{fig8:dblfld_6bands},
and then confirmed as in Fig.~\ref{fig11:EDPV_guess},
$\tilde{A}(E)$ increases sharply at 
$E_{\rm c} \pm E_{\rm w}=21.6 \pm 6.9$ keV
as if a resonance is operating.
As a conceivable resonance phenomenon in magnetars,
we speculate that this might be the proton cyclotron resonance
in the magnetic poles \citep{ProtonCycloTheory},
which will take place at an energy of
\begin{equation}
E_{\rm pcr} = 6.3 \times (B/10^{15}{\rm G}) (1+z)^{-1}~~{\rm keV}
\label{eq:PCR}
\end{equation}
where $z \sim 0.25$ is the gravitational redshift on the NS surface.
So far, spectral evidence of the proton cyclotron resonance 
has been obtained from several magnetars,
mostly as (often transient) absorption lines at typical energies of several keV
\citep{ProtonCyclo1708,Tiengo13,ProtonCycloIsolated,
ProtonCyclo0142,ProtonCyclo2259}.
Then, by setting $E_{\rm pcr}= E_{\rm c}$,
we deduce $B = 4.3\times 10^{15}$ G.
Although this is an order of magnitude higher 
than the nominal dipole MF of \oneE,  
$2.2 \times 10^{14}$ G \citep{Kuiper12},
it could be explained by the assumed local multipoles. 
In fact, it is still an order of magnitude lower 
than the toroidal field $B_{\rm t}\sim 10^{16}$ G,
suggested by the value of $\epsilon$ (Paper I).

The resonance width, $E_{\rm w}=6.9$ keV, 
may be attributed mostly to the gradients in the MF intensity,
because the thermal Doppler broadening must be much smaller,
$E_{\rm tD}\sim 0.5$ keV as judging from 
the SXC temperature (\S~\ref{subsec:spectra}).

Because the Compton cross section is inversely 
proportional  to the scatterer mass,
photon scattering by protons is usually 
negligible compared to that by electrons.
However, the electron scattering is completely inhibited 
in the present case due to the MF
(except ordinary-mode photons propagating across the MF lines),
whereas the proton scattering must be enhanced resonantly at $E \sim E_{\rm pcr}$,
by a factor $\left(E_{\rm pcr}/E_{\rm tD} \right)^2 \sim 2\times 10^{3}$,
like in electron-cyclotron resonances \citep{CRSF}.
Therefore,  photons with $E \sim E_{\rm pcr}$ would be 
resonantly scattered by protons with a cross section 
which is comparable to the Thomson cross section.
Since the NS atmosphere is usually considered Thomson thick,
a fair fraction of HXC photons with these energies, 
that are traveling with small angles to the MF lines,
would be scattered sideways by protons.
This will work as effectively increasing $\theta$,
and might explain the energy dependence of $\tilde{A}(E)$,
possibly answering the first half of [Q4].
However, admittedly, the explanations of 
$\tilde{\psi}(E)$ and $S(E)$ are yet to be found.
We might need to invoke some exotic physics,
because they both {\em break the basic time-reversal symmetry},
apparently without energy dissipation.

\subsubsection{Some thoughts on the SXC vs. HXC relation}
In the \NuS\ data,
the 36 ks pulse-phase modulation was absent at $<8$ keV,
where the SXC is dominant (Fig.~\ref{fig1:spec}).
This agrees with the results of Paper I,
and those for 4U~0142+61 \citep{Makishima14, Makishima19},
indicating that the SXC has
a symmetric emission pattern around $\hat{x}_3$.
The two spectral components are hence
inferred to differ not only in their spectral shapes,
but also in their emissivity patterns and production mechanisms.
While the HXC possibly results from the photon splitting process,
the SXC is likely to be  thermal radiation
(but modified by the strong MF) from the heated polar regions.
Since the two spectral components have comparable  luminosities,
and there is a hint of variation propagation
from the HXC to SXC (\S~\ref{subsec:light_curves}),
the SXC may be powered via two channels;
direct magnetic heating of the polar regions, and heating by the HXC.

With these in mind, let us consider [Q5],
for which we can think of at least two possibilities.
One is that the  hardest end of the SXC 
also breaks the symmetry around $\hat{x}_3$,
and the other is that the phase modulation in the HXC pulses is
enhanced at the lowest end of the HXC spectrum.
Although the former could work,
below we consider the latter scenario.
When spin flips of protons in the MF are taken into account,
the proton cyclotron resonance would take place
not only at $E_{\rm pcr}$, but also generally at \citep{ProtonCycloTheory}
\begin{equation}
E_{n, s} = (n+ s g/2)  E_{\rm pcr} 
\label{eq:PCR_harmonics}
\end{equation}
where $n$ is an integer describing Landau-level separations,
$g=5.586$ is the proton g-factor, 
and $s=0~{\rm or}~ \pm 1$ specifies the spin flips.
Equation (\ref{eq:PCR}) is the case with $n=1$ and $s=0$.
Usually, the resonance  with $s=0$ and $n=\pm 1$ has
far larger strengths than those with  $s=\pm 1$ and $n=0$,
but these conditions might change under the ultra-strong MFs.
Then, the observed resonance centre $E_{\rm c}$ 
could alternatively be identified with $E_{0,1}= 2.79 E_{\rm pcr}$,
and if so, the fundamental resonance should occur at 
$E_{\rm pcr}=21.6/2.79=7.74$ keV,
implying $B=1.5 \times 10^{15}$ G.
In this case, $\tilde{A}(E)$ will be  enhanced  at  $\sim 7.7$ keV,
and this effect might partially cancel the decreasing HCX 
contribution below $\sim 12$ keV (Fig.~\ref{fig1:spec}),
making the modulation visible down to $\sim 8$ keV.

Although the reversal in $\Psi$ is not easily explained,
its energy,  $E \sim 27$ keV, 
might fit into the scheme of equation~(\ref{eq:PCR_harmonics}).
In fact, if assuming $E_{\rm pcr}=7.74$ keV,
we expect $E_{1,1}=29.4$ keV,
which is close to the $\Psi$-reversal energy,  $E_{\rm d}$,
considering systematic uncertainties in the model form of $\tilde{\psi}(E)$.
The transition with $n=1$ and $s=\pm 1$ should be forbidden 
by the quantum selection rule at least in the 1st order perturbation,
but might be allowed in higher orders.

The above interpretation has an obvious caveat;
no spectral features (emission, absorption, or break)
are seen at any of these characteristic energies.
This issue would be our future task,
since the proton cyclotron resonance phenomenon
under  extreme MFs is currently far less  understood
than the more familiar electron cyclotron resonance
in X-ray pulsars \citep{MaxReview}.

Putting this spectral issue aside,
how about these resonances in the \Su\ observation,
when the pulse-phase modulation amplitude was
not apparently energy dependent in $10-40$ keV (Paper I)?
It could be that the multipole MF was  stronger
at that time due to the enhanced activity,
so these resonances were at $>40$ keV
where the \Su\ HXD data had  limited signal-to-noise ratio.
An alternative scenario is that the resonances 
were actually in the  10--40 keV range covered by the HXD,
but the effect was not noticeable
because $\theta$ was already rather large as considered above.
These may give an answer to the 2nd half of [Q4].

In a future work, we will return to the \Su\ observation,
and re-analyse the XIS data in the 5--12 keV range
for possible energy dependent pulse-phase behaviour.
Also, analyzing the \NuS\ data of \oneE,
acquired in 2019 for a comparable length of time
as the present data set,
is evidently our another future task.

\subsubsection{Comparison with 4U 0142+61}

To  explain the present \NuS\ data of \oneE,
we have so far developed a scenario 
which combines  such physical ideas as
the free precession of a NS that is axially deformed by
intense toroidal MFs,
the photon splitting process as the HXC emission mechanism,
the variable multipole configuration,
and the  proton cyclotron resonance.
Although the scenario is very speculative,
the invoked individual ingredients are not 
extraordinary in view of the basic physics of strong MFs,
and of the understanding of magnetars as magnetically-powered NSs.
Therefore, the scenario must be able to explain 
the behaviour of  4U~0142+61 as well.

The $T=55$ ks pulse-phase modulation in 4U~0142+61
has been detected in two out of three observations with \Su,
and one observation with \NuS\ \citep{Makishima14, Makishima19}.
In all the three cases, the effect was observed only in the HXC,
and was absent in the SXC, 
in  agreement with the behaviour of \oneE.
The modulation amplitude of 4U~0142+61
differed considerably from one observation to another;
$A/P_0 = 0.08 \pm 0.03, 0.14 \pm 0.05$, 
and $0.020 \pm 0.009$ (Table 2 of \cite{Makishima19}),
with $P_0 = 8.689$ s.
These variations in $A$ may be explained by the present scenario
as a change of $\theta$, or $a$, or both.

Since  4U~0142+61 is a much older system, 
the same mechanism as for [Q2] should explain 
why 4U~0142+61 showed on average smaller $A/P$ ratios
 than \oneE\ in 2009, and than in the present data at $E \sim E_{\rm c}$.
Likewise, 4U~0142+61 is thought to have 
its proton cyclotron resonances much below the HXC energy range.
This may explain why none of the three observations of 4U~0142+61
gave evidence of noticeable energy dependence of $A$.
Thus, the scenario is thought to apply to  4U~0142+61 as well [Q6].

Finally, the complex pulse-phase behaviour 
found in the present data would not be specific to \oneE.
We expect that similar phenomena will be revealed
by detailed hard X-ray timing studies of other magnetars,
particularly young and active ones.

\section{Conclusion}

During the \NuS\ observation made after \oneE\  
returned in quiescence,
the HXC pulses at $P_0= 2.08671$ sec
were phase-modulated with the same 36 ks period
as in the outburst observation with \Su\ (Paper I).
Because of the presence both in the outburst and quiescence,
this 36 ks periodicity can be identified with the slip period,
associated with the free precession of the NS
that is axially deformed by $\epsilon \sim 0.6 \times 10^{-4}$.
The deformation, in turn, is likely due to the internal toroidal MF 
reaching $\sim 10^{16}$ G.
The  SXC pulses were free from the phase modulation.
Therefore, the  two spectral components must be distinct 
in their emissivity patterns,
as well as in their spectral shapes.

In the present data,
the pulsed fraction was high ($\gtrsim 40\%$)  below 8 keV,
but it decreased  to $\lesssim 10\%$ in the 10--25 keV interval.
This puzzling behaviour was found to stem from strong 
energy dependences in the  HXC pulse properties,
partially coupled with the 36 ks phase modulation.
Namely, at $E\sim 22$ keV, the modulation amplitude exhibited a 
resonance-like enhancement to $\sim P_0/4$.
Over the $\sim 10$  to $\sim 27$ keV interval,
the modulation phase changed by $\sim 65^\circ$,
followed by a $\sim 180^\circ$ jump.
Regardless of the 36 ks phase, 
the overall pulse phase shifted with energy by $\sim 8\%$.
Corrections of the photon arrival times for these  effects
have successfully brought the PF to $\gtrsim 18\%$
 over the entire 8--70 keV range,
and rectified the energy-dependent irregular variations 
in the HXC pulse phase and shape.

Though still tentative and speculative,
a possible astrophysical scenario for these unexpected  results has been derived.
That is, the HXC is produced by the photon-splitting process,
and its emissivity is asymmetric around $\hat{x}_3$
due to the presence of tilted strong multipoles.
The degree of this asymmetry depends on the energy,
possibly due to a proton cyclotron resonance
which might be present at $\sim 22$ keV or $\sim 7.8$ keV,
although no spectral features are observed at these energies.
In any event, this modeling will provide a useful guideline
to future observations (including polarimetry) of this magnetar and similar objects,
and to various theoretical studies of 
physics under extreme MFs.

Our final words should be:
``The truth, however strange in itself, 
is always interesting and beautiful to seekers after it."
(quoted and modified from
Agatha Christie, {\it The Murder of Roger Ackroyd}).

\section*{Acknowledgements}
The present work was financially  supported by 
the JSPS grant-in-aid (KAKENHI),  number 18K03694.

\section*{Data availability}
The data underlying this article are available in 
the NASA/GSFC HEASARC NuSTAR Data Archive,
at  https://heasarc.gsfc.nasa.gov/docs/nustar/nustar\_archive.html


\section*{Appendix A: The $Z^2$ statistics}

Although the $Z_m^2$ method was originally proposed for  
unbinned photon-series data
\citep{Zn2_83,Zn2_94,Enoto11},
the algorithm involves 
photon-by-photon  calculations of sinusoidal functions.
The Fourier-power scheme of equation~(\ref{eq:Z2_definition})
provides an equivalent and a faster way,
and the results are insensitive to $N_{\rm bin}$ if  $N_{\rm bin} \gg m$.
In the present analysis, we employ $N_{\rm bin} =360$.

As  $m$ increases and approaches $N_{\rm bin}$,
the $Z_m^2$ evaluation becomes equivalent to the chi-square method 
as predicted by the Parseval's theorem.
In this way,  the $Z^2$ technique evaluates the pulse significance 
using only the lowest several harmonics 
which usually carry most of the signal power.
This makes the $Z_m^2$ method much less subject to the Poisson noise
than the chi-square statistics.
An additional advantage of the $Z^2$ technique is 
that it is essentially free from a natal nuisance in the other method,
namely, the choice of $N_{\rm bin}$.

For Poissonian random signals without intrinsic periodicity,
the values of $Z_m^2$ obey a 
chi-square distribution of  $2m$ degrees of freedom, 
and its probability density is given as
\begin{equation}
f(X; 2m) \propto X^{m-1} \exp(-X/2)
\label{eq:chisq_distribution}
\end{equation}
where $Z_m^2$ is abbreviated as $X$.
The distribution has the mean of  $2m$,
and the standard deviation of $2\sqrt{m}$ around it.
Then,, $Z_m^2$ is expected to increases with $m$ as
\begin{equation}
Z_{m+1}^2(P)  \sim Z_{m}^2(P) +2~.
\label{eq:Z2_scaling}
\end{equation}

When the data contain an intrinsic periodicity at $P$,
its power and the Poissonian contribution
approximately add up to make $Z_m^2$.
As a result, the increment
$\delta X \equiv X -X_0 $
%
becomes an important measure
when conducting the demodulation  in \S~\ref{sec:detailed_pulsation}.
Here,  $X_0$ denotes the value of $Z_m^2$ without demodulation.
When using  $m=1$ in particular, 
equation (\ref{eq:Z2_scaling}) for the Poissonian contribution
becomes an exponential function;
so is the upper integral of $f(X;m)$.
Therefore, a value of $X$, which is larger by $\delta X$ than a fiducial value $X_0$,
has a factor $\exp(-\delta X/2)$ 
lower chance occurrence probability than $X_0$.
This factor becomes $6.7 \times 10^{-3}$ if $\delta X = 10$.

\section*{Appendix B: The running average}

In the present work,  
pulse profiles and double-folded maps are smoothed
with a running average (RA),
where we combine three consecutive bins of   time series $\{x_n\}$ 
as $\tilde{x}_n = x_{n-1}/4+  x_n /2 +  x_{n+1}/4$,
and use $\{\tilde{x}_n \}$  in place  of $\{x_n\}$.
By suppressing high-frequency noise,
the RA  reduces the errors associated with each data bin.
If $\{x_n\}$  has an average  1-$\sigma$ error of $\delta x$, 
and if it is independent between the adjacent bins,
the 1-$\sigma$ fluctuation in $\{\tilde{x}_n \}$  becomes
$\{(1/4)^2+(1/2)^2+(1/4)^2 \}^{1/2} \delta x=  \sqrt{3/8}\; \delta x =0.61\; \delta x$.
This estimate  holds when $\{x_n\}$ varies mildly with $n$.

The above form of RA is equivalent to a Fourier filter of 
$ F(k)=\frac{1}{2} \left[ 1+ \cos(\pi k/k_0)\right]$, 
with $k$ the wave number,
and $k_0= N_{\rm bin}/2$ the Nyquist wave number.
When this filter is applied to a white-noise signal 
with the variance $(\delta x)^2 $,
the output data have a variance as (with $x \equiv k/k_0$)
\[
(\delta x)^2 \int _0^{k_0} F(k)^2  dk 
=\frac{(\delta x)^2 }{4} \int _0^{1}  \left[ 1+ \cos(\pi x) \right]^2dx  
= \frac{3}{8}(\delta x)^2 ~,
 \]
and a  standard deviation by $\sqrt{3/8}\; \delta x=0.61\; \delta x$.

We confirmed these estimates using an actual X-ray data set,
the \Su\ XIS  data of 4U~0142+61 in 2009 \citep{Makishima14}.
It consists of $N_{\rm bin}=53, 610$ bins of 2-sec counts,
with the average of $\langle x_n \rangle=44.21$ c bin$^{-1}$
and 1-$\sigma$ scatter of $\delta x = 6.97$ c bin$^{-1}$.
Since $\delta x  \approx \sqrt{\langle x_n \rangle}=6.65$,
the data can be regarded as Poisson-dominated,
although the source was pulsing, like in the present data, with a period of 8.69 s.
Through the same RA  as above,
the 1-$\sigma$ error was reduced to $\delta x=4.19$ c bin$^{-1}$,
by a factor of $4.19/6.97=0.60$ in agreement with the analytic predictions.

\section*{Appendix C: Errors associated with the $Z^2$ statistics}
Consider a data set  $\{t_{i}\} (i=0, 1, 2, .., N_{\rm tot}-1)$ 
comprising $N_{\rm tot}$ photons,
where $t_i$ denotes the arrival time of the $i$-th photon.
Let us  search the data for a periodicity at a period $P$,
using up to  the $m$-th harmonics of the Fourier power.
Usinging the  Fourier coefficients $\{a_k(P), b_k(P)\} (k=0, 1, . ., m)$
in equation~(\ref{eq:Fourier_tr}) 
which can be computed by  a Fourier transform of  $\{t_i\}$, 
we can construct a scalar quantity,
called likelihood function and denoted as $L [t_{i}; \{a_k(P), b_k(P)\}; P]$,
which expresses the probability for the data to have a periodicity at $P$.
The best-estimated period is obtained as a value of $P$
that maximizes  this likelihood.
In a more complex analysis ({\it e.g.}, the  demodulation analysis),
$L$ may depend not only on $P$, but also on some other parameters.
In this case, $L$ must be maximized 
with respect to all these  parameters.

When the time series is dominated by the Poisson noise,
there holds a relation as \citep{Yoneda_PhD}
\begin{equation}
\ln L [t_{i}; \{a_k(P), b_k(P)\}; P] \approx  \frac{1}{2} Z_m^2~.
\label{eq:Z2_vs_likelihood}
\end{equation}
Therefore,  the demodulation analysis, 
where we search for the maximum $Z_m^2$,
is equivalently to 
maximizing the  log likelihood
with respect to the four parameters, $P, A, T$, and $\psi$.

In the period search using the chi-square method,
the errors associated with the best-estimated $P$
(and other parameters) are generally difficult to estimate,
because this is suited to a  {\em minimisation} process,
and a large value of chi-square is useful only in rejecting a null hypothesis
that the signal does not have a significant periodicity at $P$.
In contrast, the $Z_m^2$ evaluation, being a maximizing process,
allows the parameter errors to be more easily formulated.

Suppose that  a solution $(P, T, A, \psi)$
gives  the maximum likelihood, $L^*$.
The true parameters may be slightly different, 
and so is  the associated $L_0$.
Again assuming the noise dominance,
the difference  $2(\ln L^* - \ln L_0)$ is known to  obey 
a chi-square distribution with $\nu$ d.o.f. \citep{Cowan}.
Here, $\nu$ is the number of parameters involved in $L$.
Combining this with equation~(\ref{eq:Z2_vs_likelihood}),
we find that the values of $Z_m^2$ around its maximum should obey,
in the present case, a chi-square distribution with $\nu=4$.
(This 4 represents  the parameter number, but  not our choice of $m=4$.)
Since its  upper 68\% probability point is 4.72, 
we define the 68\% error range of each parameter 
(with the other 3 parameters re-adjusted) 
as the point where  $Z_4^2$ decreases by 4.72 from the maximum.

\vspace*{-1mm}
\section*{Appendix D: Statistical significance of the pulse-phase modulation}
Following Paper I and \cite{Makishima19},
we evaluated the statistical probability ${\cal P}$
with which the 36 ks peak in the 8--25 keV DeMD (Fig.~\ref{fig7:Tscan_8-25keV})
appears due to chance fluctuations.
For this purpose, 
we performed the $m=4$ demodulation analysis
using the same 8--25 keV data,
but over a  range of $T=0.25-4.0$ ks,
where the pulse-phase fluctuation would not have any
enhanced periodicity,
because the range is much longer than $P_0$
but shorter than the orbital period of \NuS, 5.8 ks.
We utilised  the same scan ranges and steps
in $P$,  $A$, and $\psi$,
as in Fig.~\ref{fig7:Tscan_8-25keV}.
To  make the adjacent sampling points in $T$ mutually independent
in terms of Fourier wave numbers,
the scan step in $T$ was varied so as to satisfy
$
\Delta T \sim  T^2/T_{\rm tot}
$
where $T_{\rm tot}=151$ ks is the total observation span.
This has yielded 476  steps in $T$,
and in two cases among them, 
$Z_4^2$ exceeded the target value of 
$Z_4^2=72.95$ (Table~\ref{tbl:Z2_summary}).
Therefore, the probability of finding $Z_4$ values 
larger than was observed,
{\em at a single value of $T$}, 
is estimated as $2/476=4.2\times 10^{-3}$.

To obtain ${\cal P}$, we must multiply 
with  the effective number of trials $N_{\rm tr}$ in $T$
that  was  involved in analyzing the actual data.
We may set as  $N_{\rm tr}=1$,
because our purpose is to reconfirm the \Su\ discovery
rather than finding a new modulation period, 
and the error range of equation (\ref{eq:36ks_Suzaku}),
7.0 ks, is covered by a single Fourier wave number
which haa $\Delta T = 8.5$ ks for $T=36$ ks.
We then obtain ${\cal P} \sim 4.2\times 10^{-3}$. 
However, this could be an underestimation,
because  we may have to select $\Delta T \sim  2.1$ ks
considering  the use of $m=4$.

To avoid the above ambiguity in $N_{\rm tr}$
that arises via the use of $m>1$,
we repeated the same calculation using $m=1$ this time.
Out of the 476 trials, we  again found two cases 
(but at different values of $T$ from those found with $m=4$)
wherein $Z_1^2$ exceeded the target value of 
$Z_1^2=60.88$ (Fig.~\ref{fig7:Tscan_8-25keV}).
We hence reconfirm ${\cal P} \sim 4.2\times 10^{-3}$. 
Furthermore, a  sort of  Monte-Carlo simulation
using the actual data instead of fake data,
described in Paper I and \cite{Makishima19},
yielded 11 cases, out of 2000,
wth $Z_1^2$ higher than 60.88.
This gives ${\cal P} \sim 5.5\times 10^{-3}$.
 From these evaluations, we quote,
 as a round number, ${\cal P} = 0.5\%$.
 
\end{document}